\documentclass[aip,jcp,preprint]{revtex4-1}

\usepackage{amsmath,amssymb,bm}
\usepackage{graphicx}
\usepackage{appendix}

\begin{document}

\title{Resolving resonance effects in the theory of single particle photothermal imaging}

\author{Harrison J. Goldwyn}
\email{goldwyn@uw.edu}
\affiliation{Department of Chemistry, University of Washington, Seattle, WA 98195}
\author{Stephan Link}
\affiliation{Department of Chemistry, Rice University, Houston, TX 77005}
\affiliation{Department of Electrical and Computer Engineering, Rice University, Houston, TX 77005}
\author{David J. Masiello}
\email{masiello@uw.edu}
\affiliation{Department of Chemistry, University of Washington, Seattle, WA 98195}

\begin{abstract}
  Photothermal spectroscopy and microscopy provides a route to measure the spectral and spatial properties of individual nanoscopic absorbers, independent from scattering, extinction, and emission. The approach relies upon use of two light sources, one that resonantly excites and heats the target and its surrounding environment and a second off-resonant probe that scatters from the resulting volume of thermally modified refractive index. Over the past twenty years, considerable effort has been extended to apply photothermal methods to detect, spatially resolve, and perform absorption spectroscopy on single non-emissive molecules and other absorbers like plasmonic nanoparticles at room temperature conditions. Numerous theoretical models have been developed to interpret these experimental advances, ranging from simple theories providing transparency to a limited portion of the underlying physics to more accurate theories capable of quantitative predictions at the cost of mathematical opacity.
  Particular complications in the mathematical interpretation arise for larger target systems that host their own intrinsic scattering resonances as well as for background media that do not instantaneously thermalize with the absorbing target, problems of interest to the photothermal imaging of biological samples as well as metamaterials. 
  The aim of this Perspective is to overview the theory of photothermal spectroscopy and microscopy and present a new simplified theoretical language aimed at researchers entering the field, that makes clear the dependencies of the photothermal signal on fundamental physical parameters. This approach recovers past models in certain limits while explicitly including the effects of target scattering resonances, thermal and optical retardation, and lock-in detection that have not yet been incorporated into such a model. Focus is made on plasmonic particles to interpret the photothermal signal, yet all results are applicable equally to individual molecules or nanoparticle absorbers. Consequently, we expect this Perspective to provide a useful foundation for the understanding of photothermal measurements independent of target identity.
\end{abstract}

\maketitle



\section{Introduction}
From Hooke and Leeuwenhoek's first glimpse of microorganisms in the 1600s \cite{oldy,micrographia} until today, optical microscopy has provided a fundamental characterization tool of the microscale. Optical microscopy has endured hundreds of years of technological advancement, owing to the remarkable signal to noise, biocompatibility, and molecular specificity of fluorescence \cite{Malte2013fluo}. More recently, with the development of robust single molecule detection \cite{orrit1990single,keller1996single,moerner1999illuminating} and super-resolution imaging \cite{sahl2013super,hess2006ultra,betzig2006imaging,sharonov2006wide,deschout2014precisely}, optical microscopy has even been extended to operate below the diffraction limit, resolving the details of individual molecules and their interactions with their environment \cite{moerner1999illuminating}. However, despite the success of such fluorescence-based imaging techniques, most absorbing molecules do not efficiently fluoresce, but instead convert optical excitation into heat through nonradiative downconversion. This reason alone has compelled researchers to develop alternative methods to characterize the microscopic world based solely upon absorption.

Detecting not only the appearance of photons but also their disappearance would provide a complete understanding of how a nanoscale object processes light. But isolating absorption from emission and scattering is difficult. The first absorption based single molecule detection was demonstrated in 1989 at liquid helium temperatures, where molecular absorption cross sections are $10^6$ times larger than at ambient conditions due to quick thermal dephasing \cite{moerner1989optical}. Performing such measurements at room temperature would require large laser intensities, introducing the problem of detecting the depletion of a few photons from many. Even with ultrasensitive photodetectors, this kind of measurement would be difficult to achieve since background scattering by interfaces and an inhomogeneous environment can conflate true absorption within an extinction measurement \cite{kukura2010single,celebrano2011single}. To isolate absorption from a nanoscale object like a single molecule, a direct measurement may not be ideal.

\begin{figure}
\includegraphics[width=\textwidth]{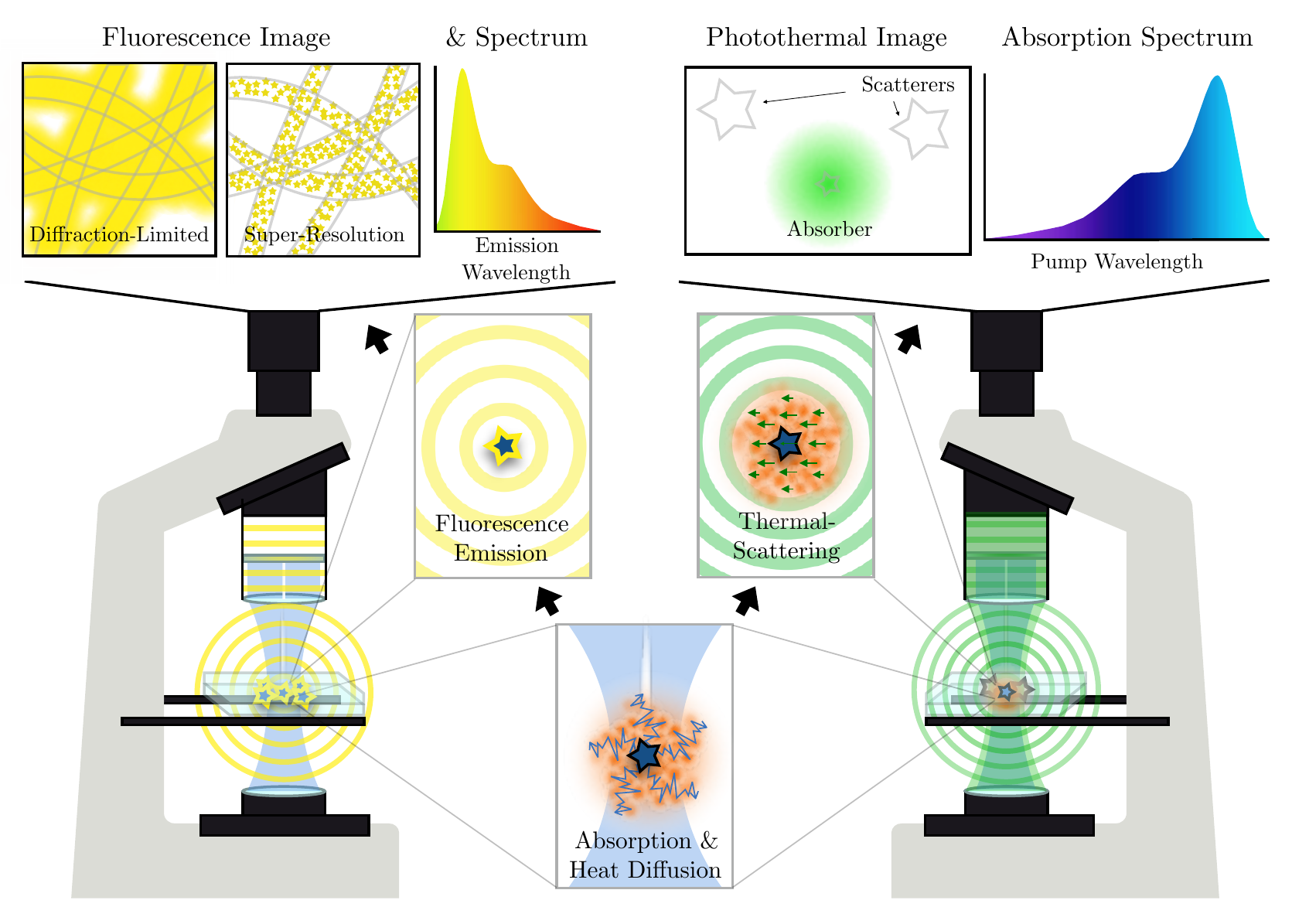}
\caption{ \label{fig:1}
    \fontsize{12pt}{12pt}\selectfont
    Illustrated comparison of fluorescence imaging and spectroscopy (left) with photothermal imaging and photothermal absorption spectroscopy (right), the latter being capable of optically detecting individual non-fluorescent absorbing particles. In fluorescence microscopy or spectroscopy, the sample is illuminated with light tuned to fluorescence absorption (blue). The fluorescent molecules absorb and later emit at a Stokes shifted wavelength (yellow). This emission is collected through a filter that blocks excitation light and is either directed towards a camera to form an image or into a spectrometer to form a fluorescence emission spectrum. In the single molecule limit, the molecular point spread function can be fit to form a sub-diffraction-limited map of single molecule locations, known as a super-resolution image \cite{sahl2013super,hess2006ultra,betzig2006imaging,sharonov2006wide,deschout2014precisely}. With the isolated emission from each molecule, single molecule spectra are also obtainable \cite{ambrose1999single,moerner2003methods,lee2019spectral,doi:10.1021/acs.jpclett.9b02270}. In the photothermal measurement, the same absorption process occurs. But in this case the heat dissipated by the absorbing molecule or particle is used as a ``thermal lens''. A second probe beam (green) is focused onto and then scatters off of this heated region surrounding the absorber due to slight variations in the refractive index with temperature. Locking in to modulation of the absorbed pump beam (undepicted) allows for detection of individual molecules \cite{gaiduk2010room} with minimal background from non-absorbing scatterers \cite{Boyer1160}. Single particle absorption spectra can also be measured by tuning the pump wavelength \cite{yorulmaz2015single}. 
    }
\end{figure}

Photothermal imaging overcomes the difficulties of detecting individual nanoscale absorbers by measuring absorption second hand \cite{tokeshi2001determination,Boyer1160,cognet2003single,berciaud2004photothermal}. The technique relies on lock-in detection to changes in scattering of a probe laser by an object with oscillating absorption of a second pump laser. This is accomplished by amplitude modulating the pump laser, which induces an oscillating ``thermal lens'' around the absorber as heat diffuses into the environment. Based on this approach together with a some variations, detection \cite{gaiduk2010room,adhikari2020photothermal} and even spectroscopy \cite{yorulmaz2015single,berciaud2007absorption,berciaud2005observation,berciaud2005photothermal,joplin2017correlated,joplin2017imaging,yorulmaz2016absorption,adhikari2020photothermal,knapper2016chip,heylman2016optical,barnes1994femtojoule,barnes1994photothermal,larsen2011ultrasensitive,larsen2013photothermal,yamada2013photothermal,chien2018single,doi:10.1021/acs.nanolett.9b02796,nanolett17,li2015absorption,Aleshire2288} of single particle absorption at room temperature is now possible. Fig.\ \ref{fig:1} illustrates the principles underlying photothermal imaging in comparison with traditional fluorescence imaging.  

Due to their large absorption efficiencies, plasmonic nanoparticles make excellent nanoheaters \cite{Kraemer:2011aa,ni2016steam} and have been commonly employed test cases for photothermal detection. Soon after the development of photothermal imaging, the demonstration of high signal-to-noise detection of nanometer-sized gold and silver nanoparticles was reported with corresponding theory that explained potential optimization of photothermal signal \cite{Boyer1160,berciaud2004photothermal,berciaud2006photothermal}. The theoretical trends in signal amplitude and spatial resolution with beam-aperture, heating beam intensity, and modulation frequency \cite{berciaud2006photothermal} demonstrated good agreement with experiment while maintaining relative simplicity, setting the standard for theoretical interpretations of the measurement. But over time, alternative models and associated interpretations have been presented \cite{selmke2012photothermal} that explain the measured data equally well. Simultaneously, interest has also developed in the photothermal imaging of larger plasmonic particles with non-negligible scattering \cite{bhattacharjee2019active}, a feature missing from the original models aimed at detecting single molecule absorption. Even for the metal nanoparticles employed as photothermal imaging test targets, accounting for scattering is unnecessary for absorbers that are small in comparison to the wavelength of light. Taken together, the apparent differences in photothermal models as well as their omission of resonant scattering effects highlight the need for an improved understanding of photothermal imaging theory.

It is the purpose of this Perspective to provide a simple and robust mathematical language that reveals the complexity of the photothermal signal of resonant scatterers through a general theoretical formalism. Although our perspective is aimed at researchers new to the field, it affirms the use of previously published interpretations in certain physical limits. We show below that by deriving the photothermal signal from fundamental assumptions about lock-in detection and accounting for all fields at plat, we arrive at the starting assumptions of other researchers in different limits: some working in the limit of negligible pure-scattering contribution and others working in the limit of fast thermal diffusion. 
While the qualitative understanding of these limits is not new, we present the explicit mathematical assumptions behind them. Beyond this understanding, the influence of the target's scattering resonances upon the photothermal signal represents an important new direction of inquiry that may be exploited to enhance signal \cite{shi2020photothermal,li2020resonant} or locate windows of spectral transparency to perform photothermal experiments in the absence of scattering resonance effects. We extent our analytic theory to this regime and examine the qualitative trends that can be seen explicit in the resulting equations. We begin by first reviewing the theoretical background of photothermal imaging, highlighting the foundational principles commonly used in the literature. Next we derive the photothermal image explicitly accounting for the lock-in detection process and investigate all possible contributing mechanisms of probe scattering such as the effects of thermal retardation and target scattering resonances. Finally, a set of detailed examples are presented to illustrate the effects of the background medium thermal and optical properties as well as single-particle resonances upon the measured photothermal image.

\section{Theoretical Background}

Significant advances in the analysis of the photothermal signal were made in 2010 by Orrit and coworkers \cite{gaiduk2010detection} with the aim of pushing detection to the limit of a single molecule. Their analysis expanded upon earlier work with Lounis and coworkers \cite{Boyer1160}, and examined the trend in photothermal signal-to-noise with the thermo-optical properties of the thermally conductive environment. Theory and experiment both showed a linear trend in photothermal signal with the so-called {\it photothermal strength} of the background medium (proportional to the product of the inverse thermal conductivity $\kappa$ and thermo-optic coefficient $dn/dT$). Link and coworkers later demonstrated that this analysis could be used to push the sensitivity of photothermal imaging even further by immersing the absorbers in thermotropic liquid crystals \cite{chang2012enhancing}. This study extended the theoretical-experimental agreement on the trend in signal-to-noise with background photothermal strength significantly, as the liquid crystal used had approximately three times the photothermal strength of the the strongest medium studied by Orrit, a path they have continued to innovate on \cite{ding2016hundreds}. 

Despite these theoretical advances in modeling the photothermal signal, explicit treatment of the lock-in detection is often omitted. Instead, the signal is taken to be proportional to the interference of the oscillating component of the scattered probe field $\mathbf{E}_\Omega$ with the transmitted/reflected (depending on experimental geometry) probe field $\mathbf{E}^\mathrm{tr/re}_\mathrm{pr}$ observed in absence of the target sample. Written in terms of these complex fields with assumed harmonic time dependence, the photothermal image power takes the form \cite{berciaud2006photothermal}
\begin{equation} \label{eq:I_PT_Lounis}
P_\mathrm{PT}^{(1)} \propto \mathrm{Re}[ \mathbf{E}^\mathrm{tr/re}_\mathrm{pr} \cdot \mathbf{E}_\Omega^*]
\end{equation}
with the proportionality constant depending only upon basic material and incident light field parameters.

Independently, three distinct yet complementary theoretical constructions of the photothermal signal were published by Cichos and coworkers in 2012: a generalized Lorenz-Mie theory to solve the scattering problem rigorously including effects of aberration on focused laser beams \cite{selmke2012photothermal}, a simplified Fraunhofer diffraction approach yielding analytical expressions \cite{Selmke:12}, and a ray-optics treatment taking advantage Gaussian matrix optics \cite{selmke2012gaussian}. 
The relative predictive power and qualitative differences of these and past models has already been reviewed \cite{selmke2015physics}. But for our purposes it is worth noting a foundational assumption not obviously equivalent to Eq.\ \ref{eq:I_PT_Lounis}.
Selmke \emph{et al}.\ start with photothermal signal defined as the difference in transmitted/reflected probe light with the pump beam on and off. Said differently, the photothermal image is the difference between images of the heated and room temp systems,
\begin{equation} \label{eq:I_PT_Cichos}
P_\mathrm{PT}^{(2)} \propto P_\mathrm{hot} - P_\mathrm{room}.
\end{equation}
The assumption that the modulated experiment can be modeled by this static theory was validated phenomenologically by comparing demodulated images to their steady state equivalents \cite{selmke2012photothermal}, with good agreement found between the theory and experiment. 
%
%
This definition of the post lock-in photothermal signal as a different measurement has since become ubiquitous. And with the development of iSCAT microscopy, \cite{ortega2012interferometric,lindfors2004detection} the language of photothermal imaging theory has become more transparent \cite{shi2020photothermal,adhikari2020photothermal,li2020resonant}. Each of the two terms in Eq.\ \ref{eq:I_PT_Cichos} can be interpreted as an iSCAT measurement of the heated and cold system separately.
iSCAT is simply a scattering measurement performed under conditions to observe not only the scattered intensity of the target but also the interference between the scattered field and the reflected incident (or reference) field. The measurement can also be performed in transmission, in which case it has been termed COBRI \cite{huang2021quantitative}). 
In the reflection geometry, the observed intensity is $I_\mathrm{iSCAT} \propto |E_\mathrm{refl} + E_\mathrm{scatt}|^2 = |E_\mathrm{refl}|^2 + |E_\mathrm{scatt}|^2 + 2\mathrm{Re}[E_\mathrm{refl} \cdot E_\mathrm{scatt}^*]$. For small particles the scattered field will scale with the particle volume $V$, while the pure scattering contribution $|E_\mathrm{scatt}|^2 \propto V^2$ will be relatively negligible. 
With the iSCAT signal written in this form its utility in detection of small particles is clear. Rewriting Eq.\ \ref{eq:I_PT_Cichos} for the photothermal signal as the difference of iSCAT measurements of the heated and room temp targets yields
\begin{equation} \label{eq:I_PT_Cichos_iSCAT}
P_\mathrm{PT}^{(2)} \propto |E_\mathrm{scatt}^\mathrm{hot}|^2 - |E_\mathrm{scatt}^\mathrm{cold}|^2 + 2\mathrm{Re}[E_\mathrm{refl}^* \cdot (E_\mathrm{scatt}^\mathrm{hot} - E_\mathrm{scatt}^\mathrm{cold})]
\;.
\end{equation}
In what follows, we will show through explicit derivation of the lock-in detection that Eq.\ \ref{eq:I_PT_Cichos_iSCAT} is equal to Eq.\ \ref{eq:I_PT_Lounis} if we neglect both the pure scattering terms and any time dependence of heat diffusion. 

With the ever increasing capabilities of lithographically engineered nanostructures, interest has arisen in controlling the nanoscale temperature distribution within absorbing nanoclusters composed of plasmonic particles. In 2019, Willets, Link, Masiello and coworkers adapted photothermal imaging toward quantifying the temperature distribution within and around interacting plasmonic nanoclusters as a form of optical thermometry \cite{bhattacharjee2019active}. Because detection was not the primary goal, these systems are larger than the common subjects of photothermal imaging, with dimensions on the order of the pump wavelength. Consequently their scattering cross sections exceed absorption, yet the effect of metal scattering on the photothermal image was at that point unestablished. Regardless, reasonable agreement was found between experiment and simulated photothermal images by considering only the pure scattering contribution to Eq.\ \ref{eq:I_PT_Cichos_iSCAT} as well as accounting only for the effect of the metal nanostructure heating directly and not the thermal lens, despite the literature precedent being the opposite for smaller particles. This was later confirmed through direct simulation of both pure scattering and interference contributions to Eq.\ \ref{eq:I_PT_Cichos_iSCAT} in similar systems \cite{jebeli2021wavelengthdependent}, serving as evidence that the plasmonic target had reached a new size regime where the metal scattering was the dominant contribution to the photothermal signal and not scattering through heated background matrix. More recent theoretical work has acknowledged that as nanoparticle size increases, the room temperature scattering contributes to the photothermal signal even after lock-in detection \cite{shi2020photothermal}. These effects can be greatly enhanced near the particles' scattering resonances, which \citet{li2020resonant} demonstrated through fitting a dipolar model of the Eq.\ \ref{eq:I_PT_Cichos_iSCAT} to experimental observations of the trend in photothermal signal with particle resonance wavelength. Continuing to build understanding of the photothermal imaging of resonant nanoparticles is also important to biological applications where the background scattering can be highly heterogeneous \cite{nedosekin2012photothermal,he2015label,he2016noninvasive}. In the domain where the pure scattering $\propto |E_\mathrm{scatt}^\mathrm{hot}|^2 - |E_\mathrm{scatt}^\mathrm{cold}|^2$ dominates, we avoid difficulties of the heterogeneous phase sensitivity of the interference term between reflected/transmitted incident beam and the scattered field $\mathrm{Re}[E_\mathrm{refl}^* \cdot (E_\mathrm{scatt}^\mathrm{hot} - E_\mathrm{scatt}^\mathrm{cold})]$.

In this Perspective, we seek to illuminate past assumptions behind models of the photothermal signal with a simple theory aimed at researchers new to the field. While maintaining a transparent mathematical formalism, we will explore regimes where the target hosts it own unique scattering resonances as well as when it does not. The former is the regime of larger plasmonic nanoparticles then have been studied most commonly in the context of photothermal imaging, which will be the target sample of choice henceforth.  However, the theory presented here can equally be applied to individual molecular absorbers/scatterers with appropriate modification of the target polarizability. To accomplish these objectives we present a simple analytic theory that extends to larger particle sizes where scattering from the target's dipole resonance contributes to the photothermal signal. Following an explicit derivation of the lock-in measurement and careful accounting of the participating electric fields, we show that the photothermal signal reduces to Eq.\ (\ref{eq:I_PT_Lounis}) in the limit of small scattering and (\ref{eq:I_PT_Cichos}) in thermally-static limit. We then investigate the physical differences between the two terms comprising the most general expression for the photothermal signal and discuss the experimental and physical factors controlling their relative contribution to the total signal. This discussion is most clear in the thermally-static limit, which is developed in greatest detail and followed with a comprehensive discussion of the more accurate thermally-retarded model that is used for calculation of all presented Figures. The model predictions are compared to previously published data on the trend in photothermal signal with background medium and then extended to larger particles. We observe probe wavelength dependence that is clearly connected to the particle scattering cross sections and background medium. Finally we assess the qualitative difference in photothermal images produced by Eqs.\ (\ref{eq:I_PT_Lounis}) and (\ref{eq:I_PT_Cichos}) in the context of past work.

\section{Outline of the model: Lock-in detected photothermal signal}

\begin{figure}
\includegraphics[width=\textwidth]{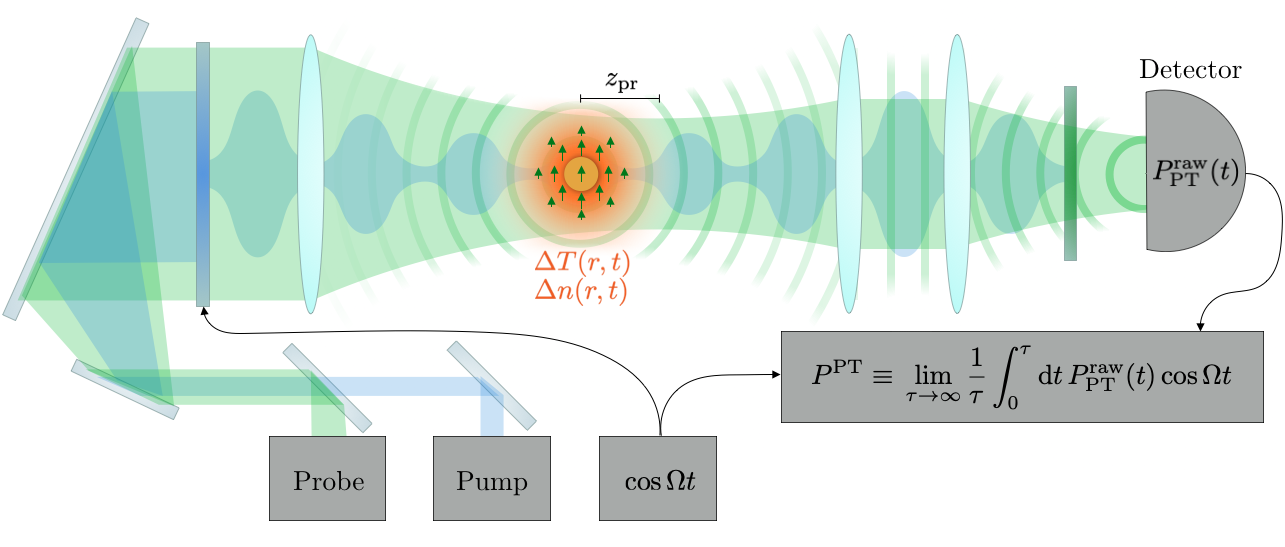}
\caption{ \label{fig:2}
    Illustration of the photothermal imaging setup as modeled. The pump beam (blue) passes through a modulating filter and is focused onto a  plasmonic gold nanoparticle. The nanoparticle absorbs the pump beam and acts as an oscillatory source of heat. The heat diffuses and generates the elevated temperature distribution $\Delta T({\bf x}, t)$. The continuous wave probe beam (green) is also focused nearby the nanoparticle with offset $z_\mathrm{pr}$ along the optical axis that can drastically impact the phase of the signal. The probe field scatters not only off of the nanoparticle, but also the space and time varying refractive index perturbation $\Delta n({\bf x}, t)$ of the background medium, resulting from the local temperature increase. The scattered and transmitted probe light is focused onto a photodetector. This raw photothermal signal $P_\mathrm{PT}^\mathrm{raw}$ is fed to a lock-in amplifier. The lock-in integrates the raw photothermal signal against the modulation reference signal $\cos\Omega{t}$ (neglecting any phase offset, discussed in text) to filter out static background from the photothermally induced scattering.}
\end{figure}
The lock-in detected photothermal signal can be understood by first working through a simplified version of the model, but the general form of the photothermal signal we arrive at is robust to accounting for both metal and thermal-lens effects, as well as the finite speeds of both thermal diffusion (see SI Section S2) and light (see SI Section S3). A diagram of the experiment is shown in Fig.\ \ref{fig:2}. As in the rest of this Perspective, we will assume two monochromatic lasers are focused near an isolated absorbing nanoparticle. We distinguish the lasers as the pump and probe beams with intensities in the focal spot $I_\mathrm{pump}$ and $I_\mathrm{probe}$ with corresponding optical frequencies $\omega_\mathrm{pu}$ and $\omega_\mathrm{pr}$. The pump beam is amplitude modulated and the probe beam is continuous. The probe beam is scattered by the sample, which is collected in either a reflection or transmission (diagrammed) geometry. Although the experimental geometry does affect some subtle aspects of the analysis, we will leave an investigation of those differences to a future work and focus of the the common physics. The probe-frequency light collected at the detector is demodulated electronically in order to select the photothermal signal produced by temperature induced changes in scattering, which should have the time dependence as the pump beam modulation. We will derived this demodulated signal with explicit accounting of all relevant fields and discuss the approximations that arise in turn to preserve closed mathematical forms. 

In what follows, we will break the model into distinct physical steps: (\ref{subsection:absorption}) the absorption of the modulated pump beam by the nanoparticle, (\ref{subsection:heating}) the heating of the nanoparticle and surrounding background medium, (\ref{subsection:scattering}) scattering of the continuous wave probe beam by the heated nanoparticle, and (\ref{subsection:detection}) detection and demodulation of the photothermal signal to extract the temperature induced scattering. Gaussian units are used throughout.

\subsection{Absorption} \label{subsection:absorption}

The pump beam is intensity modulated at frequency $\Omega$ $(\sim\mathrm{kHz})$ much slower then optical frequencies, therefore the time averaged intensity absorbed by the particle will still be slowly varying in time,
\begin{align}
{I}_\text{pump}(t; \omega_\mathrm{pu})
={}&
I_\mathrm{pu} \cdot\frac{1}{2}\left(1+\cos\Omega t\right)
\\
={}& \label{eq:I_pump}
\frac{cn}{8\pi}\left|\mathbf{E}_\mathrm{pu} e^{-i \omega_\mathrm{pu} t}\right|^2 \cdot\frac{1}{2}\left(1+\cos\Omega t\right),
\end{align} 
where $I_\mathrm{pu}\equiv({cn}/{8\pi})\left|\mathbf{E}_\mathrm{pu} e^{-i \omega_\mathrm{pu} t}\right|^2$ is the plane wave intensity pre-modulation. Although Eq.\ \ref{eq:I_pump} is an ansatz consistent with previous literature \cite{berciaud2006photothermal,gaiduk2010detection,selmke2015physics}, the associated beating electric field is given in 
SI Eq.\ (S41)
. The spatial dependence of the beam is left implicit within the field amplitude $\mathbf{E}_\mathrm{pu}$. 

Due to the timescale argument above, the nanoparticle continuously absorbs the wave $I_\mathrm{pu}$ with overall scaling $(1/2)\left(1+\cos\Omega t\right)$. The power absorbed by the nanoparticle located at point $\mathbf{x}_0$ is therefore
\begin{equation}
P_\text{abs}(t) 
= 
\sigma_\text{abs}(\omega_\mathrm{pu}) I_\mathrm{pu}\cdot\frac{1}{2}(1+\cos\Omega t),
\end{equation}
where variation in the field across the absorbing nanoparticle at location $\mathbf{x}_0$ has been neglected. We see that the effect of the modulated pump beam is simply to modulate the power absorbed due to their linear relationship under the given assumptions. In the following, the absorption cross section of the nanoparticle target will be modeled in the long wavelength limit, so that it can be described by its dipole polarizability.

\subsection{Heating} \label{subsection:heating}

For simplicity, we will now make a second timescale argument. Let us assume that the nanoparticle heats up and reaches thermal equilibrium with the surrounding medium much faster than the modulation time $2\pi/\Omega$. The system can then be considered to quasistatically oscillate between hot and room temperature states, with the temperature following the steady state heat-diffusion equation. This approximation that speed of heat diffusion is effectivly infinite in relaxed in SI Section S2, and does not affect the general form of the photothermal signal derived below. For a spherical particle absorbing power $P_\text{abs}$, we assume all of this energy is converted to heat sufficiently quickly to equate the modulated absorbed power with the heat flux leaving the sphere's surface. With the radius $a$ and isotropic background thermal conductivity $\kappa_b \ll \kappa_\text{metal}$, the sphere is heated to the uniform temperature
\begin{equation}
  \begin{split}
\Delta T_\text{NP}(t) 
&\equiv T_\text{NP}(t) - T_R \\
&= \frac{P_\text{abs}(t)}{4\pi\kappa_b a} \\
&= \label{eq:static_temp}
 \langle T_\mathrm{NP} \rangle (1+\cos\Omega{t}),
\end{split}
\end{equation}
where the average nanoparticle temperature across a modulation cycle is defined by
\begin{equation} \label{eq:average_temp}
\langle T_\mathrm{NP} \rangle = \frac{1}{2}\frac{\sigma_\text{abs}(\omega_\mathrm{pu}) I_\mathrm{pu}}{4\pi\kappa_b a}
\end{equation}
and $T_R$ is the ambient (room) temperature in the absence of the pump. For other shapes the denominator varies, but the linear scaling with power absorbed is maintained \cite{baffou2013thermo}, and this analysis proceeds as general.

The major simplification of the thermally static limit evoked above is that the temperature of the medium surrounding the particle oscillates with the same time dependence as the particle. Specifically, the temperature outside the sphere will follow
\begin{equation}\label{eq:steady_state_gen_temp}
\Delta T_b(\mathbf{x}, t) \equiv T_b(\mathbf{x}, t) - T_R = \frac{a}{|\mathbf{x}|} \Delta T_\text{NP}(t).
\end{equation}
At the elevated temperatures attainable in experiment, the refractive indices for both medium and metal (both real and imaginary parts) increase linearly with temperature as
\begin{equation}
n(T) 
= 
n(T_R) + \frac{{d}n}{{d}T}\Bigg|_{T_R} 
\!\!\!\!\!
\cdot
(T(\mathbf{x}, t) - T_R).
\end{equation}
Therefore, the thermally modulated refractive index profile will inherit the same time dependence as the temperature.

\subsection{Scattering} \label{subsection:scattering}

To detect this refractive index fluctuation, a probe beam illuminates the sample throughout the heating modulation. The probe is a continuous wave optical laser with field in the focal region described by $\mathbf{E}_\mathrm{pr} e^{-i \omega_\mathrm{pr} t}$. As with the pump beam, the spatial dependence of the probe beam profile is implicit in $\mathbf{E}_\mathrm{pr}.$ This field is scattered by the nanoparticle and surrounding thermal lens at beat frequencies $\omega_\mathrm{pr} \pm \Omega$ as the hot and cold system scatter differently. As for the absorption process, we make the long wavelength approximation, allowing us to treat the probe beam scattering off of metal nanoparticle and its surrounding heated region of oscillating refractive index as sourced by an induced electric dipole $\mathbf{p}(t) = \boldsymbol{\alpha}(t) \cdot \mathbf{E}_\mathrm{pr}$, with polarizability defined by the nanoparticle temperature but accounting for polarization of the heated background surrounding the nanoparticle as well. Note here that the scattering dipole oscillates at two separate timescales, first with the kHz modulation frequency contained within the temperature $T_\mathrm{NP}$, and second with the optical probe frequency which we may leave implicit as it will average out trivially below. In the above and in what follows, any explicit function of time $f(t)$ will refer to modulation time-scale variation. 

Cichos and coworkers have discussed the limitations of this point dipole model \cite{Selmke:12,selmke2012photothermal,selmke2012gaussian}, however, it is our aim here only to present the simplest model of the photothermal signal (see Ref.\ \cite{selmke2015physics} for specific quantitative limitations of this dipole approach). Continuing with the model construction, if the change in all refractive indices is small compared to the room temperature indices, i.e., $({{d}n}/{{d}T}|_{T_R}) / n \ll 1$, then the polarizability (accounting for the dipole response of both metal and heated background) can be Taylor expanded and truncated at first order as
\begin{equation} \label{eq:temp_induced_alpha}
\alpha(t) 
\approx
\alpha_{T_R}
+ 
\left.
\frac{
	{d} \alpha
	}
	{
	{d} n 
	} 
\frac{
	{d} n 
	}
	{
	{d}T
	}\right|_{T_R} 
\!\! 
(T_\mathrm{NP}(t)-T_R)
.
\end{equation}
Note that while the polarizability model is still left unspecified, its linear response inherits time dependence directly from $T_\mathrm{NP}(t)-T_R \propto (1+\cos\Omega{t})/2$. We have also left the refractive index $n$ general, so that the current derivation applies to simple \emph{background ball} model of thermal lensing presented in Section IVA as well as the core-shell model of metal and coarsely discretized thermal lens discussed in SI Section S3 and used for all computations.

The scattered field sourced by the induced dipole $\mathbf{p}(t)$ separates into the time-independent and time-dependent contributions
\begin{equation}
\begin{split}
\mathbf{E}_\mathrm{scat}(\mathbf{x}, t) 
&=
\left[ 
    \mathbf{E}_0(\mathbf{x})  
    +  
    \mathbf{E}_\Omega(\mathbf{x}) \cos{\Omega}t
    \right] e^{-i \omega t}
,\end{split}
\end{equation}
following the definition of the temperature induced polarizability in Eq.\ \ref{eq:temp_induced_alpha} and the linear response $\mathbf{E}_\mathrm{scat}(\mathbf{x}, t) = \mathbf{G} (\mathbf{x},\mathbf{x}_0) \cdot \boldsymbol{\alpha}(t) \cdot \mathbf{E}_\text{probe}(t)$, where $\mathbf{G} (\mathbf{x},\mathbf{x}_0)$ is the standard dipole relay tensor in the far-field limit \cite{jackson2007classical}. 
The time-independent scattered field can be written in terms of the polarizability components as
\begin{equation} \label{eq:static_field}
\begin{split}
\mathbf{E}_0 (\mathbf{x})
&\equiv
\mathbf{G} (\mathbf{x},\mathbf{x}_0)
\cdot
\pmb{\alpha}_0
\cdot 
\mathbf{E}_\mathrm{pr}(\mathbf{x}_0)
\\
&=
    \mathbf{G}(\mathbf{x},\mathbf{x}_0) \cdot \left(
    \boldsymbol{\alpha}_{T_R}
    + 
    \left.
    \frac{
        {d} \boldsymbol{\alpha}
        }
        {
        {d} n 
        } 
    \frac{
        {d} n 
        }
        {
        {d}T
        }\right|_{T_R} 
        \langle T_\mathrm{NP} \rangle 
    \right)
    \cdot \mathbf{E}_\mathrm{pr}(\mathbf{x}_0),
\end{split}
\end{equation} 
which contains the sum of the room temperature scattering and a static offset resulting from the amplitude modulation of the particle temperature oscillating about some value greater than room temperature.
The time-dependent or slowly-varying component of the scatted field reads
\begin{equation}\label{eq:modulated_field}
\begin{split}
\mathbf{E}_\Omega(\mathbf{x})\cos\Omega{t}
&\equiv
\mathbf{G} (\mathbf{x},\mathbf{x}_0)
\cdot
\pmb{\alpha}_\Omega(t)
\cdot 
\mathbf{E}_\mathrm{pr}(\mathbf{x}_0)
\\
&=
\mathbf{G}(\mathbf{x},\mathbf{x}_0)
\cdot \left(
    \left.
    \frac{
        {d} \boldsymbol{\alpha}
        }
        {
        {d} n 
        } 
    \frac{
        {d} n 
        }
        {
        {d}T
        }\right|_{T_R} 
        \langle T_\mathrm{NP} \rangle 
        \cos\Omega{t}
        \right)
    \cdot \mathbf{E}_\mathrm{pr}(\mathbf{x}_0),
\end{split}
\end{equation} 
arising from the temperature-induced variation in the local refractive index. 

Both scattered fields will increase in complexity upon addition of thermal retardation and the resonant scattering (see SI Sections S2-S3), but the separation of the field into static and oscillating components remains. In this case, the contribution of the nanoparticle temperature to the static scattered field will also have interesting consequences on the photothermal signal.

\subsection{Detection} \label{subsection:detection}

After frequency filtering the the pump beam out of the detection path, the light focused onto the photodetector can be interpreted as the superposition of the transmitted/reflected probe field and the scattered probe field. This raw (pre-lock-in) photothermal intensity arriving at the detector can be represented generally by 
\begin{equation}
I_\mathrm{PT}^\mathrm{raw}(\mathbf{x}, t)
= 
\frac{cn}{8\pi} \left|
    \mathbf{E}^\mathrm{tr/re}_\mathrm{pr}(\mathbf{x})
    +
    \mathbf{E}_\mathrm{scat}(\mathbf{x},t)
    \right|^2
\;,\end{equation} 
but this is not exactly representative of the experimentally measured photothermal signal. The common experimental setup involves measuring a single voltage reported by a photodiode, which is then fed to the lock-in amplifier to filter out unwanted background. In this case, it is more accurate to consider the integrated intensity profile at the detector to be the raw photothermal signal $P_\mathrm{PT}^\mathrm{raw} (t) = \int_{D}I_\mathrm{PT}^\mathrm{raw}(\mathbf{r}_D,t){d}A$ that will be converted to a voltage by the photodiode. We can calculate this integration at the detector by making another simplifying assumption. Given that the scattered fields have already been approximated as electric dipole radiation, we assume that the transmitted/reflected probe field differs negligibly in spatial distribution. With both fields assumed to be dipolar, we employ conservation of energy to perform the integration on the scattered intensity (before focusing) across a spherical surface spanning a solid angle defined by the collection objective. As discussed in 
SI Section S1
in the context of modeling the magnitude of the transmitted/reflected probe field, the spatial form of the intensity is now $I_\mathrm{PT}^\mathrm{raw}(\mathbf{r}_D)= I_\mathrm{PT}^\mathrm{raw} ({\hat{\bf p}^2 - ({\hat{\bf r}_D\cdot\hat{\bf p}})^2})/(k_\mathrm{pr}r_D)^2,$ where $\mathbf{r}_D = r_D\hat{\bf r}_D \equiv \mathbf{x}_D - \mathbf{x}_0$ is the vector connecting the nanoparticle position to the observation point in the integration domain and the wave vector magnitude for all collected fields is $k_\mathrm{pr} = \omega_\mathrm{pr} n/ c$. Integration over the spatial dependence of the collected intensity simply scales the entire signal as a function of the collection angle $\theta_\mathrm{col}$. It will therefore be useful to employ the result of the detection integral $P_\mathrm{PT}^\mathrm{raw}(t) = I_\mathrm{PT}^\mathrm{raw} f(\theta_\mathrm{col}) / k_\mathrm{pr}^2$, to introduce a scalar notation for the collected fields $E_i$ without their spatial dependence, defined by $\mathbf{E}_i(\mathbf{r}_D) = {E}_i ({\hat{\bf p}-\hat{\bf r}_D\hat{\bf r}_D\cdot\hat{\bf p}}) e^{i k_\mathrm{pr} r_D}/(k_\mathrm{pr} r_D)$. The functional form of $f(\theta_\mathrm{col})$ is derived in 
SI Section S1.

A photothermal image produced by this technique is analogous to confocal microscopy, generated by rastering the sample (nanoparticle at point $\mathbf{x}_0$) through the beam path (focused onto point $\mathbf{x}_f$) and recording the photothermal signal as a function of sample position. This procedure is elaborated on in Section \ref{section:image}, but the spatial dependence of the rastered image arises implicitly through the spatial profile of the probe beam $\mathbf{E}_\mathrm{pr}(\mathbf{x}_0;\mathbf{x}_f)$ in Eqs.\ (\ref{eq:static_field}) and (\ref{eq:modulated_field}) as well as the pump beam spatial profile within the temperature dependent polarizability.  In principle the intensity profile of the focused image at the detector $I_\mathrm{PT}^\mathrm{raw}({\mathbf{r}_D})$ could be recorded directly by replacing the photodiode with a lock-in camera, analogous to a wide-field microscope image. Although wide-field photothermal microscopy is of interest \cite{huang2021quantitative}, we will leave a detailed treatment of the latter case behind for now and focus on the confocal geometry, but most results presented below are straightforward to translate to wide-field measurement.

To model the demodulation of the signal experimentally accomplished with a lock-in amplifier, we integrate the raw photothermal signal arriving at the photodetector against the reference signal with tunable phase $\cos(\Omega{t} - \phi)$. In the idealized thermally-static limit we have enforced in this example, the phase difference between the raw photothermal signal and the lock-in reference signal affects only the overall magnitude of the post-lock-in signal since all time dependence of the scattered field is proportional to ${E}_\Omega(t) = {E}_\Omega \cos{\Omega}t$. We will in SI Section S3 that thermal retardation effects introduce different phase components to the raw photothermal signal and the choice of $\phi$ will be significant. But in the thermally-static limit we can compute the lock-in detected photothermal signal as follows,  
\begin{equation}
\begin{split}
P^\text{PT} 
&\equiv
\lim_{\tau\to\infty} \frac{1}{\tau}\int_0^\tau  {d}t\,  P_\mathrm{PT}^\mathrm{raw}  \cos(\Omega t)\\
&= 
\frac{c n}{4\pi}
\mathrm{Re}\!\left[
    {{E}_\mathrm{pr}^\mathrm{tr}}\cdot {E}_\Omega^* 
    + 
    {E}_0 \cdot {E}_\Omega^*  
    \right]
\frac{f(\theta_\mathrm{col})}{k^2}\\
&\equiv P^\mathrm{PT}_\mathrm{PI} + P^\mathrm{PT}_\mathrm{SI}.
\end{split}
\end{equation}
The photothermal signal consists of two terms of different physical origin. The first, recognizable from Eq.\ \eqref{eq:I_PT_Lounis}, describes the interference between the transmitted/reflected probe field and the slowly-varying component of the scattered field. We will refer to it as the \emph{probe-interference} term, $P^\mathrm{PT}_\mathrm{PI} \equiv({c n}/{4\pi}) \mathrm{Re}\left[{{E}_\mathrm{pr}^\mathrm{tr}}\cdot {E}_\Omega^* \right] f(\theta_\mathrm{col})/k_\mathrm{pr}^2$. The second describes the interference of the static and slowly-varying components of the scattered field, or the \emph{self-interference} of the scattered probe field, $P^\mathrm{PT}_\mathrm{SI} \equiv({c n}/{4\pi})\mathrm{Re}\left[{{E}_0}\cdot {E}_\Omega^* \right] f(\theta_\mathrm{col})/k_\mathrm{pr}^2$. 

As discussed above, this feature of the photothermal signal comprised of probe- and self-interference is a general result. But in the thermally static limit we may rewrite both terms to better compare to models of the signal used elsewhere in the literature and discussed above in Eqs.\ \eqref{eq:I_PT_Lounis}--\eqref{eq:I_PT_Cichos_iSCAT}. 
By definition of the fields $E_0$ and $E_\Omega$ in Eqs.\ \eqref{eq:static_field}--\eqref{eq:modulated_field}, we can relate $E_0$ and $E_\Omega$ to the scattered fields by an analogous set of completely static experiments: one with the pump beam continuously driving the system at $I_\mathrm{pu}$ (the maximum intensity reached in Eq.\ \eqref{eq:I_pump} over a modulation cycle), and the second with the pump beam removed and the system at room temperature. The scattered fields associated with these \emph{hot} and \emph{cold} experiments are
\begin{equation}
\begin{split}
{E}_\mathrm{scat}^\mathrm{hot}
&= 
{E}_0 + {E}_\Omega 
\\
{E}_\mathrm{scat}^\mathrm{cold}
&= {E}_0 - {E}_\Omega.
\end{split}
\end{equation}
Although maybe unintuitive, the above relations result from the the time-independent field ${E}_0$ inheriting some temperature dependence from the static offset in the intensity modulation, the $1$ in the $(1 + \cos\Omega{t})$. Substituting these relations into the probe- and self-interference terms yields
\begin{align}
P^\mathrm{PT}_\mathrm{PI}
&= \label{eq:thermaly_static_PI}
\frac{c n}{4\pi}
\mathrm{Re}\!\left[
    {{E}_\mathrm{pr}^\mathrm{tr}}\cdot \left(
        {E}_\mathrm{scat}^\mathrm{hot}
        -
        {E}_\mathrm{scat}^\mathrm{cold}
        \right)^* 
    \right]
\frac{f(\theta_\mathrm{col})}{k^2}\\
P^\mathrm{PT}_\mathrm{SI}
&= \label{eq:thermaly_static_SI}
\frac{c n}{8\pi}\left[
    |{E}_\mathrm{scat}^\mathrm{hot}|^2 - |{E}_\mathrm{scat}^\mathrm{cold}|^2  
    \right] \frac{f(\theta_\mathrm{col})}{k_\mathrm{pr}^2},
\end{align}
which are exactly the components of the photothermal signal posited in Eq.\ \eqref{eq:I_PT_Cichos}.

What we have found is two fold. First result we just discussed, the general photothermal signal reduces to that in Eq.\ \eqref{eq:I_PT_Cichos_iSCAT} commonly used by researched in the thermally-static limit. The second major result is that the thermally-static Photothermal signal decomposed into the form of Eqs.\ \eqref{eq:thermaly_static_PI}--\eqref{eq:thermaly_static_SI} is effectivly the difference in two iSCAT measurements, as discussed around Eq.\ \eqref{eq:I_PT_Cichos_iSCAT}. From here we can see that it is the neglect of the pure scattering contribution that gets us to the photothermal signal posited in Eq.\ \eqref{eq:I_PT_Lounis}, but more generally the complete neglect of the self-interference or the limit ${{E}_\mathrm{pr}^\mathrm{tr}}\gg {E}_0$. In light of these results together with the past literature, it is natural to ask: what physical parameter regimes separate these two limits of the photothermal signal? We will attempt to answer this and related questions moving forward through rigorous modeling of the various scattering processes that contribute to the observed signal.

\section{Models and mechanisms of probe scattering} 

Here we develop a model of photothermal scattering from the heated background medium surrounding the target nanoparticle only. Both thermal retardation and effects of the metal on scattering are neglected for simplicity (but later accounted for in 
SI Sections S2--S3
). This approach, along with its deviations from the more accurate version of the theory that follows, will reveal the physical origin of the photothermal signal in various parameter regimes. Following literature precedent, we will examine the dependence of the photothermal signal on the thermo-optical properties of the background medium and compare results to data from Refs. \cite{gaiduk2010detection,chang2012enhancing}. We will then eliminate these two approximations and assess the impact. In all models, the spatial profile of the temperature-induced refractive index perturbation in the background medium is averaged over. Therefore, some aspects of the thermal lensing discussed in Refs. \cite{selmke2012photothermal,Selmke:12,selmke2012gaussian} are not captured. Nevertheless, what the point-dipole model loses in quantitative accuracy of the spatial distribution of scattered light is made up for in the transparency of the role of its parameters.

\subsection{Background ball model} \label{section:background_ball}

Before moving forward with a model of the scattering process that accounts for both the target nanoparticle and surrounding thermal lens, we will first follow the literature precedent and proceed under the assumption that the nanoparticle can be neglected from the scattering process and serves only as a heater to generate the thermal lens. The photothermal signal is therefore created by scattering only from the heated region of background material surrounding the nanoparticle, typically glycerol. We will first model the scattering polarizability as an isotropic sphere of heated background medium immersed in an infinite room temperature thermal background, and refer to this scenario as the \emph{background ball} model. Since the polarizability is the same in all three space directions, it reduces to the scalar polarizability $\alpha.$

The sphere of heated background material has no room temperature polarizability, but does respond linearly to temperature increases. Accounting for optical retardation effects as discussed in 
SI Section S3
, the polarizability varies with temperature according to 
\begin{equation}
\left.
\frac{{d}{\alpha}}{{d}n_b}
\right|_{T_R}
=
\frac{2}{3} b^3 \frac{1}{n_b(2q_s - 1)^2},
\end{equation}
defined by the wavelength-dependent retardation factor $q_s = ({1}/{3})[1 - (kb)^2- i({2}/{3}) (kb)^3]$, the wave vector magnitude $k=\omega_\mathrm{pr} n_b/c$, the ball radius $b$, and the background index at room temperature $n_b$. To define the temperature of this heated background ball, we assign it the volume averaged temperature of the thermal lens. Carrying out the volume integral of Eq.\ \ref{eq:steady_state_gen_temp}, the average temperature of this ball is related to the nanoparticle temperature by
\begin{equation}
\Delta T_\mathrm{gly} = \frac{3}{2}\frac{f^2-1}{f^3-1}\Delta T_\mathrm{NP},
\end{equation}
where $f = (a/b)^3$ is the volume ratio of the metal absorber to the background-ball. The background-ball radius is set equal to the cross-sectional radius of the probe focal spot to represent the optically active region of heated background.

The scattered field can now be constructed from this heated background ball. Remembering that the spatial dependence of the dipole field has already been integrated out, dipole relay tensor post integration is 
defined by ${G}(\mathbf{x}_D,\mathbf{x}_0) = G (\mathbf{1}-\hat{\bf r}_D\hat{\bf r}_D) e^{i k_\mathrm{pr} r_D}/(k_\mathrm{pr} r_D)$ and therefore $G = k_\mathrm{pr}^3$. The oscillating scattered field takes the form
\begin{equation}
\begin{split} 
{E}_\Omega
&=
{G}
\cdot 
\left(
    \left.
    \frac{
        {d} {\alpha}
        }
        {
        {d} n_b 
        } 
    \frac{
        {d} n_b 
        }
        {
        {d}T_\mathrm{NP}
        }\right|_{T_R} 
        \langle T_\mathrm{NP} \rangle 
    \right)
    \cdot {E}_\mathrm{pr}(\mathbf{x}_0)
\\
&= \label{eq:bball_E0_presort}
    \left(\frac{\omega_\mathrm{pr}n_b}{c}\right)^3
 \cdot \left(
    \frac{2}{3} b^3 \frac{1}{n_b(2q_s - 1)^2}
    \frac{3}{2}\frac{f^2-1}{f^3-1}
    \left.
    \frac{
        {d} n_b 
        }
        {
        {d}T
        }\right|_{T_R} 
        \!\!\!\!
        \langle T_\mathrm{NP} \rangle 
    \right)
    \cdot 
\frac{4}{w_\mathrm{pr}} \sqrt{\frac{P_\mathrm{pr}}{ {cn_b} }}\;.
\end{split}
\end{equation}
The probe field magnitude in the focal spot is defined assuming the nanoparticle lies at the focus of a Gaussian beam with waist $w_\mathrm{pr}$ determined by the numerical aperture of the illumination objective $\mathrm{NA}_\mathrm{ill}$ (see 
SI Section S4
).

Before sorting terms in Eq.\ \eqref{eq:bball_E0_presort}, it is instructional to note the source of different physical factors. Specifically, we will analyze the dependence of the photothermal signal on the background refractive index. In past work the refractive index dependence of this expression has been assumed to follow $n_b({{d} n_b}/{{d}T}|_{T_R})$, which would arise by taking the factor of $n_b^2$ out of ${G}$ and combining with the polarizability. Here we find that if the incident probe field is defined by the laser power, it should also have background dependence. In addition, the time-averaged nanoparticle temperature 
\begin{equation}
\langle T_\mathrm{NP} \rangle 
=
\frac{1}{2}
\frac{
    I_\mathrm{pu}
    }{
    4\pi\kappa_b a
    }
4\pi \frac{\omega_\mathrm{pu} n_b}{c}
\mathrm{Im}\left[\alpha^\mathrm{metal}_\mathrm{sphere}(\omega_\mathrm{pu})\right]
\end{equation}
contains background dependence both explicitly from the a factor of the wave vector magnitude $k = \omega n_b / c$ that comes with the definition of the absorption cross section and implicitly in the polarizability. Assuming the pump intensity is also defined by the beam power, it should not have $n_b$ dependence as the prefactor will cancel with $n_b^{-1/2}$ dependence of the pump field (see 
SI Section S4
).

Separating the geometry- and frequency-dependence from the thermo-optical properties of the background (outside the absorbing polarizability), 
\begin{equation} 
{E}_\Omega
=\label{eq:bball_mod_field}
K_\Omega(\omega_i, a, b)
L_\Omega(P_i, w_i)
M(n_b, \frac{{d}n_b}{{d}T}, \kappa)
\mathrm{Im}\left[\alpha^\mathrm{metal}_\mathrm{sphere}(\omega_\mathrm{pu})\right],
\end{equation}
where the spectral and geometric parameters are contained within the retardation index $K_\Omega(\omega_i, a, b)$, the pump/probe power and focal waist are contained within the laser index $L_\Omega(P_i, w_i)$, and the thermo-optical properties of the background medium are contained within the material index $M_\mathrm{PI}(n_b, {{d}n}/{{d}T}, \kappa)$ that we will find is inherited by the probe-interference. Explicitly, 
\begin{align}
K_\Omega(\omega_i, a, b) ={}& \label{eq:K_Omega}
    \left(
        \frac{b^3}{a}
        \frac{1}{(2q_s - 1)^2}
        \frac{f^2-1}{f^3-1}
        \frac{\omega_\mathrm{pr}^3\omega_\mathrm{pu}}{c^4}
        \right)
\\
L_\Omega(P_i, w_i) ={}& 
    4\sqrt{\frac{\pi}{c}} \sqrt{\frac{P_\mathrm{pr}}{\pi w_\mathrm{pr}^2 }}
    \frac{P_\mathrm{pu}}{ \pi w_\mathrm{pu}^2}
\\
M_\mathrm{PI}(n_b, \frac{{d}n}{{d}T}, \kappa) ={}& \label{eq:MPI}
    \left(
        \frac{n_b^{5/2}}{\kappa_b}
        \left.
        \frac{
            {d} n_b 
            }
            {
            {d}T
            }\right|_{T_R} 
        \right).
\end{align}

\begin{figure} 
\includegraphics{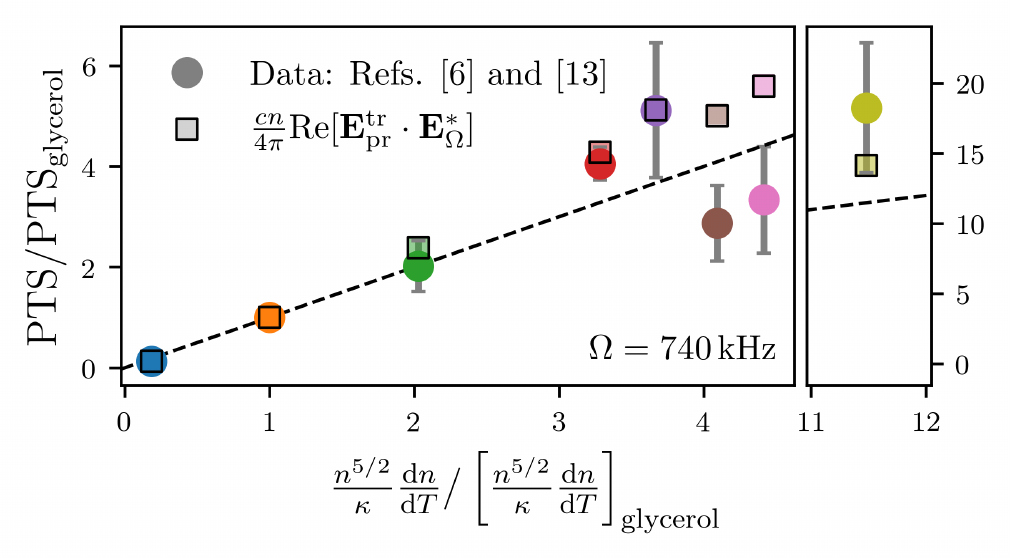}
\caption{ \label{fig:3}
    Analytical model accounting for thermal-retardation and metal scattering qualitatively predicts experimentally observed trend in photothermal signal (PTS) with background medium for gold nanoparticles of radius $a = 10$ nm. The model and data are plotted against the background material thermo-optical properties as they appear in the thermally-static model neglecting metal scattering ($M_\mathrm{PI}$, see Eq.\ \eqref{eq:MPI}). The model contains no free parameters, but both model and data are normalized to their values in glycerol. The data is from Refs. \cite{gaiduk2010detection} and \cite{chang2012enhancing} (experiment performed in reflection geometry) and background media are water (blue), glycerol (orange), ethanol (green), hexane (red), dichloromethane (pink), pentane (purple), chloroform (brown), and the liquid crystal 5CB (yellow). The photothermal signal from gold nanoparticles of radius $a = 10$ nm generally increases with the background material index ($x$-axis). Deviations of the model from the line of slope 1 (dashed black) are primarily caused by changes in the absorption cross section with $n_b$, and the effect of the glass substrate on the absorption and scattering steps. Better agreement between model and data would require a quantitative model of (1) the reflection coefficient determining the reflected probe field magnitude and (2) the effect of the glass as a heat sink in the heat diffusion step of the model. Preliminary evidence for these factors leading to the nonlinear shape of the data is described in SI Section S7. Parameters used for calculation are as reported is Ref. \cite{gaiduk2010detection}, with pump wavelength $\lambda_\mathrm{pu} = 514$ nm, pump power $P_\mathrm{pu} = 270$ $\mu$W, probe wavelength $\lambda_\mathrm{pr} = 800$ nm, probe power $P_\mathrm{pr} = 23$ mW, $\mathrm{NA}_\mathrm{ill} = \mathrm{NA}_\mathrm{col} = 1.4$, $\Omega = 740$ kHz. The probe focus was $z_\mathrm{pr} = 50$ nm from the particle. 
    }
\end{figure}

Combing this expression for the modulated scattered field with a model for the transmitted probe field discussed in Eqs.\ 
(S7, S8)
, we can write the probe-interference component of the locked-in photothermal signal defined by $P^\mathrm{PT}_\mathrm{PI} \equiv({c n_b}/{4\pi}) \mathrm{Re}\left[{{E}_\mathrm{pr}^\mathrm{tr}}\cdot {E}_\Omega^* \right] f(\theta_\mathrm{col})/k_\mathrm{pr}^2$ as,
\begin{align}
P^\mathrm{PT}_\mathrm{PI}
={}&
\mathrm{Re}[K_\mathrm{PI}(\omega_i, a, b)]
L_\mathrm{PI}(P_i, w_i)
M_\mathrm{PI}(n_b, \frac{{d}n_b}{{d}T}, \kappa)
\mathrm{Im}\left[\alpha^\mathrm{metal}_\mathrm{sphere}(\omega_\mathrm{pu})\right] 
\;,\end{align}
with modified retardation and laser indices
\begin{align}
K_\mathrm{PI}(\omega_i, a, b) 
={}& 
\label{eq:K_PI}
    K_\Omega(\omega_i, a, b) \frac{c}{\omega_\mathrm{pr}}
\\
L_\mathrm{PI}(P_i, w_i) 
={}& \label{eq:PI_term_L_factor}
\sqrt{
    \frac{3 P_\mathrm{pr}}{cf(\theta_\mathrm{col})}
    }
L_\Omega(P_i, w_i) 
f(\theta_\mathrm{col})
=
\sqrt{3 f(\theta_\mathrm{col})}
 \frac{P_\mathrm{pr}}{\pi w_\mathrm{pr} }
    \frac{P_\mathrm{pu}}{ \pi w_\mathrm{pu}^2},
\end{align}
where $\theta_\mathrm{col} = \sin^{-1}(\mathrm{NA}_{\mathrm{col}}/n_b)$. Not only do we see the well-known bi-linear power dependence of the photothermal signal \cite{gaiduk2010detection,selmke2015physics}, but it is now clear how the photothermal signal varies with background medium, assuming that the probe-interference term dominates and the change in absorption cross section with background is negligible. This assumption is consistent with the first expression for the photothermal signal posited in Eq.\ \eqref{eq:I_PT_Lounis}, and its accuracy is tested in Fig.~\ref{fig:3} by plotting the linear trend in $P^\mathrm{PT}_\mathrm{PI}$ with $M_\mathrm{PI}$ against experimental data from Refs. \cite{gaiduk2010detection,chang2012enhancing} (colored circles with error bars) as well as the more accurate model developed below that accounts for metal scattering, and the finite speeds of thermal diffusion and light (colored squares). For the $a=10$ nm particles investigated in those studies and the appropriate experimental parameters, the self-interference term is completely negligible. Despite good agreement between the linear background-ball model and the data, the data presents some variation away from the linear trend that our model recuperates after incorporation of metal scattering and thermal- and electrodynamic-retardation. Some of this can be accounted for by acknowledging change in the absorption cross section with background medium not captured in $M_\mathrm{PI}$, and more so by adding the effects of thermal retardation and metal scattering to the model as we will discuss in the following section.  

For all calculations presented here, the absorbing metal polarizability has been taken to be the exact dipolar Mie polarizability (See SI Section S5). The main effect of changing the refractive index on the absorbing polarizability is to red-shift the resonance with increasing $n_b$. This would affect the photothermal signal if the resonance shifts onto or off of resonance with background medium change. We consider pump wavelengths near the interband transitions to avoid this effect.

In the context of this background-ball model it is especially convenient to compare the probe- and self-interference terms, pointing to why the self-interference is negligible in the Fig.~\ref{fig:3} example. In this case the static and slowly-varying scattered fields are equivalent except for an overall $\cos\Omega{t}$, therefore ${E}_\Omega = {E}_0$. The self-interference is proportional to the complex magnitude squared of the field in Eq.\ \eqref{eq:bball_E0_presort},
\begin{align}
P^\mathrm{PT}_\mathrm{SI}
={}& \label{eq:SI_term_intermsof_factors}
|K_\mathrm{PI}(\omega_i, a, b)|^2
L_\mathrm{SI}(P_i, w_i)
M_\mathrm{SI}(n_b, \frac{{d}n_b}{{d}T}, \kappa)
\mathrm{Im}\!\left[\alpha^\mathrm{metal}_\mathrm{sphere}(\omega_\mathrm{pu})\right]^{\!2}
\;,
\end{align}
which again separates into retardation, laser, and material indices. The probe-interference retardation index shows up again here but squared, while the self-interference laser and material indices are
\begin{align}
L_\mathrm{SI}(P_i, w_i)
={}& \label{eq:SI_term_L_factor}
4f(\theta_\mathrm{col})
\frac{P_\mathrm{pr}}{\pi w_\mathrm{pr}^2 }
\left(\frac{P_\mathrm{pu}}{ \pi w_\mathrm{pu}^2}
\right)^2
\;,
\\
M_\mathrm{SI}(n_b, \frac{{d}n_b}{{d}T}, \kappa)
={}& \label{eq:SI_term_M_factor}
\left(
    \frac{n_b^{2}}{\kappa_b}
    \left.
    \frac{
        {d} n_b 
        }
        {
        {d}T
        }\right|_{T_R} 
    \right)^{\!2}.
\end{align}
Both probe- and self-interference terms scale linearly with the probe power, but the self-interference scales quadratically with the pump power. This means that in absence of experimental constrains (like detection limits at low power and nanoparticle melting at high power), the relative contribution of $P^\mathrm{PT}_\mathrm{PI}$ and $P^\mathrm{PT}_\mathrm{SI}$ could be tuned by the pump power. Alternatively, we can interpret the different pump power dependence of the probe- and self- interference as linear vs. quadratic scaling with the average nanoparticle temperature $\langle T_\mathrm{NP} \rangle$. We note here that although Eq.\ \eqref{eq:SI_term_L_factor} may seem to contradict the well known bi-linear power dependence of the photothermal signal, we see in Fig.\ \ref{fig:3} that the self-interference term was completely negligible. We will see below that the self-interference can arise in larger particles \cite{shi2020photothermal,li2020resonant} but most studies have not yet probed this regime. The self-interference also scales differently with the background thermo-optical properties, therefore the self-interference will increase in relative contribution for background media with larger refractive optical constants and smaller thermal constants. 

Another apparent difference between probe- and self-interference terms is their dependence on the numerical aperture of the illumination and collection objectives. In the reflection setup, there is only one objective, but in the transmission setup $\mathrm{NA}_\mathrm{ill}$ and $\mathrm{NA}_\mathrm{col}$ can be different. In the model presented, the illumination objective determines the beam waist of the probe and pump beams. For sake of example, we assume here the near diffraction limited resolution $\mathrm{FWHM}_i=0.61({\lambda_i}/{\mathrm{NA}_\mathrm{illu}})$. The Gaussian beam width is related to the FWHM by $w_i = {\mathrm{FWHM}_i}/{\sqrt{2\ln{2}}}$, which in terms of the wavelength and numerical aperture is 
\begin{equation}
w_i = 0.52\frac{\lambda_i}{\mathrm{NA}_\mathrm{illu}},
\end{equation}
for $i$ being the pump or probe. The effect of the collection objective is visible in the influence of $\theta_\mathrm{col}$ within Eqs.\ \eqref{eq:PI_term_L_factor} and \eqref{eq:SI_term_L_factor}. The difference between the two results because the magnitude of the transmitted/reflected probe beam picks up its own dependence on $\theta_\mathrm{col}$ when defined by the total transmitted/reflected power.

One last difference of note between the probe- and self-interference is that the probe field is linear in the reflection/transmission coefficient of the experimental geometry. If reflection is dominated by the glass/background interface inside the sample, then the reflection coefficient is $0.02$ to $0.2$ times the transmission coefficient. Therefore when switching from the transmission to the reflection geometry of the experiment, the probe-interference will decrease by 1-2 orders of magnitude relative to the self-interference term. This result indicates that a detailed accounting of the type of reference field may be necessary to build a quantitative model of a specific photothermal experiment. We hope that our mathematical formalism may serve as a starting point for the analysis of similar measurements to the parameters sets investigated here.

\subsection{Thermal retardation}

Although the derivation is somewhat more involved, it adds little complexity to the final result to include the full time dependence of the heat diffusion equation describing the time-dynamics of the the induced thermal field. We will leave the details to 
SI Section S2
and discuss the implications here. The solution to the steady-state heat diffusion equation in Eq.\ \eqref{eq:static_temp} is replaced with the solution to the time-dependent heat diffusion equation around a sphere of oscillating heat source density. This generalization accounts for the finite speed of heat diffusion through the background medium, or in our context the finite modulation frequency that characterizes the system's time dependence. The resulting temperature distribution around the spherical particle in Eq.\ \eqref{eq:static_temp}  can be averaged across the probe beam focal spot just as before to define an effective temperature of the optically active region of heated background.

The main difference from the thermally static theory presented above is that the temperature, and therefore the time-dependent scattered field, is no longer proportional to $\cos{\Omega{t}}$. Instead the temperature describes the superposition of four different thermal waves propagating away from the nanoparticle. All of them have the same frequency $\Omega$ and propagate through to the pre-lock-in photothermal signal, but only one is in phase with the heat source $\propto \cos\Omega{t}$. The lock-in now plays the additional role of selecting a single phase component from the modulated signal. We therefore generalize the post-lock-in photothermal signal to $P^\text{PT}(\phi) \equiv \lim_{\tau\to\infty} \frac{1}{\tau}\int_0^\tau {d}t\, I_\mathrm{PT}^\mathrm{raw} (t) \cos(\Omega t - \phi)$ which contains a tunable phase.  In reality, the experimental setup may introduce unwanted phase delays along various beam paths and by electronic components. The phase is therefore selected to maximize the signal-to-noise, which we also do for all calculations herein.

One complexity worth noting from the introduction of thermal retardation is that the connection between the self-interference term and the difference in static hot and room-temperature images described in Eq.\ \eqref{eq:thermaly_static_SI} is lost. But in all calculations performed here, despite using the thermally-retarded solutions, we did not find results qualitatively different from the thermally static limit. This can justified by the large modulation frequency (or all time variation slow) but an simple quantifier of this regime change is the thermal radius $r_\mathrm{th} = \sqrt{{2\kappa}/{C_p\Omega}}$, which characterizes the length scale of thermal diffusion away from the nanoheater in terms of previously defined parameters and the background thermal conductivity per volume $C_p$. If the thermal radius approaches the optically active volume of heated background (or probe beam radius $w_\mathrm{pr}$), than the effects of thermal retardation will be non-negligible \cite{baffou2013thermo}. But for the modulation frequency used in Figs.\ \ref{fig:4}-\ref{fig:7}, $\Omega = 100$ kHz, and $r_\mathrm{th}\approx 1499$ nm in glycerol and 1237 nm in pentane, so $r_\mathrm{th} \gg b$.

\subsection{Metal scattering}

As the target nanoparticle size increases beyond $a\sim20$ nm, the metal becomes a larger fraction of the optically active volume. Even off resonance, the change in the metal refractive index with temperature can be comparable to that of the background medium background medium and it is clear that the metal could affect the probe scattering and resulting photothermal signal. Again leaving a detailed discussion to 
SI Section S3
, scattering of the heated metal nanoparticle and surrounding background can be described analytically by the polarizability of a spherical core-shell system. Expanding this polarizability to first order in the temperature of the nanoparticle and temperature of the local background (treated again as a volume average over the spatial temperature distribution) results in
\begin{equation} \label{eq:alphacs_of_TsTc}
\alpha_\mathrm{cs} (T)
=
\alpha(T_R)
+
\left.\frac{{d}{\alpha}}{{d}n_c}
\frac{{d}n_c }{{d}T} \right|_{T_R}
        \Delta T_\mathrm{core}
+
\left.\frac{{d}{\alpha}}{{d}n_s}
\frac{{d}n_s }{{d}T} \right|_{T_R}
    \langle \Delta T(r, t) \rangle_{V_\mathrm{shell}}.
\end{equation}
The room temperature polarizability $\alpha(T_R)$ is that of the metal absorber. The linear terms describe the change in polarizability with the complex refractive index of the metal core $n_c$ as well as the refractive index of the heated background shell $n_s$. The expansion coefficients ${{d}{\alpha}}/{{d}n_c}$ and ${{d}{\alpha}}/{{d}n_s}$ are written explicitly in 
SI Section S3
for both the electrostatic (Eqs.\ 
(S26--S28)
)
and retardation corrected (Eqs.\
(S30--S32)
)
regimes. Note that although the shell consists of the same material as the background medium, we have introduced the new index $n_s$ to differentiate the temperature dependent shell from the room temperature background.
The nanoparticle temperature can be defined as before in either the thermally static or retarded limits, by evaluating the temperature distribution outside the particle at the sphere surface. With these expansion coefficients defined, the full photothermal signal can be developed following the same procedure as for the background-ball model in the previous section. This with the thermal retardation correction is the model used for calculations in all figures.

\begin{figure} 
\includegraphics[width=\textwidth]{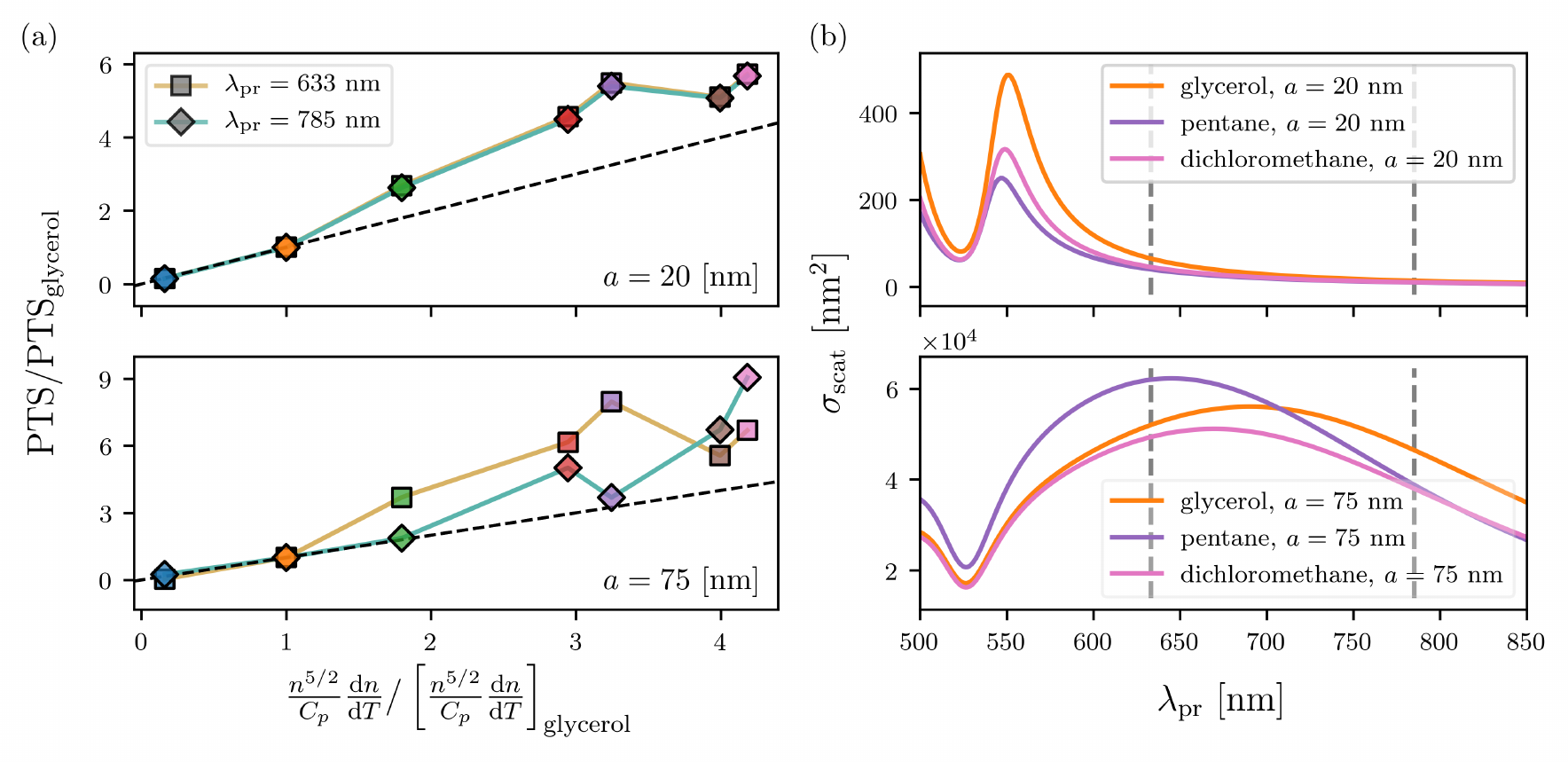} 
\caption{ \label{fig:4}
    Photothermal signal trend with background medium (a) changes with probe wavelength for bigger particles (bottom row), but not for smaller particles (top row). This effect is not explainable by simply looking at the room temperature nanoparticle scattering on the glass substrate (b). The two probe wavelengths chosen (633 and 785 nm) are overlaid on the scattering cross sections (vertical dashed-gray lines). 
    For the $a=20$ nm particle the trend in (a) is frequency independent for the chose parameters, which is consistent with the minor change in scattering cross section across glycerol, pentane, and dichloromethane (as examples). But for the $a=75$ nm particle the trend is unpredictable by the associated scattering cross sections. 
    A deeper analysis of the photothermal resonance is needed to understand the observed $\lambda_\mathrm{pr}$ dependence in trend with background medium (a). Parameters used for calculation were chosen from Ref. \cite{chang2012enhancing}, with pump wavelength $\lambda_\mathrm{pu} = 532$ nm, pump power $P_\mathrm{pu} = 25$ $\mu$W, probe power $P_\mathrm{pr} = 250$ $\mu$W, $\mathrm{NA}_\mathrm{ill} = 1.4$, $\mathrm{NA}_\mathrm{col} = 0.7$, $\Omega = 100$ kHz, and $z_\mathrm{pr}=50$ nm. Background media corresponding to the colored data points are water (blue), glycerol (orange), ethanol (green), hexane (red), pentane (purple), dichloromethane (pink), chloroform (brown), and the liquid crystal 5CB (yellow).
    }
\end{figure}

The main effect of the metal scattering is that the plasmon resonance affects the probe scattering in a way that is non-trivially related to the particle's room temperature cross section. Fig.\ \ref{fig:4} demonstrates the evolution of the trend in photothermal signal with background medium as the particle size is increased. Different from Fig.\ \ref{fig:3}, the thermally retarded equivalent of the material index is used, replacing $\kappa$ with $C_p$. This is analogous to the photothermal strength coined in past publications with additional factors of $n_b$ we believe to be correct. In Fig.\ \ref{fig:4}a, a slightly larger sphere of radius $a=20$ nm shows no variation in photothermal signal trend with background between two example probe wavelengths at the chosen particle--probe-focus offset of $z_\mathrm{pr}=50$ nm. This may be expected from looking at the scattering cross sections (Fig.\ \ref{fig:4}b) in glycerol (orange), pentane (purple), and dichloromethane (pink), which sees the strongest photothermal signal enhancement of the given backgrounds. In either case, the two chosen probe wavelengths (dashed-gray) are off-resonance. But for a much larger sphere of radius $a=75$ (Fig.\ \ref{fig:4}a, bottom) where the scattering peak overlaps both $\lambda_\mathrm{pr}$, the photothermal signal trend with background is significantly different at the two wavelengths. It is tempting to explain this with just the room temperature cross sections as an indicator of increased metal scattering with sphere size. For the $a=75$ nm particle at $\lambda_\mathrm{pr}=633$ nm the scattering trends are consistent, going from glycerol to pentane increases scattering while going from pentane to dichloromethane reduces the scattering. But at $\lambda_\mathrm{pr}=785$ nm the correspondence is broken. Despite similar scattering in pentane and dichloromethane, we see a stark difference in the relative photothermal signal. We note that the trends presented in Fig.\ \ref{fig:4}a change drastically with $z_\mathrm{pr}$ (see SI Fig S7) \cite{HWANG2007487}. Since the probe-interference term picks up an extra phase of $k_\mathrm{pr}z_\mathrm{pr}$, increasing $z_\mathrm{pr}$ introduces $\lambda_\mathrm{pr}$ dependence even for the $a=20$ nm particle. With these complexities considered, it is clear that the photothermal signal depends on the metal's resonance in a non-trivial way.

\section{The photothermal resonance}

\begin{figure} 
\includegraphics[width=\textwidth]{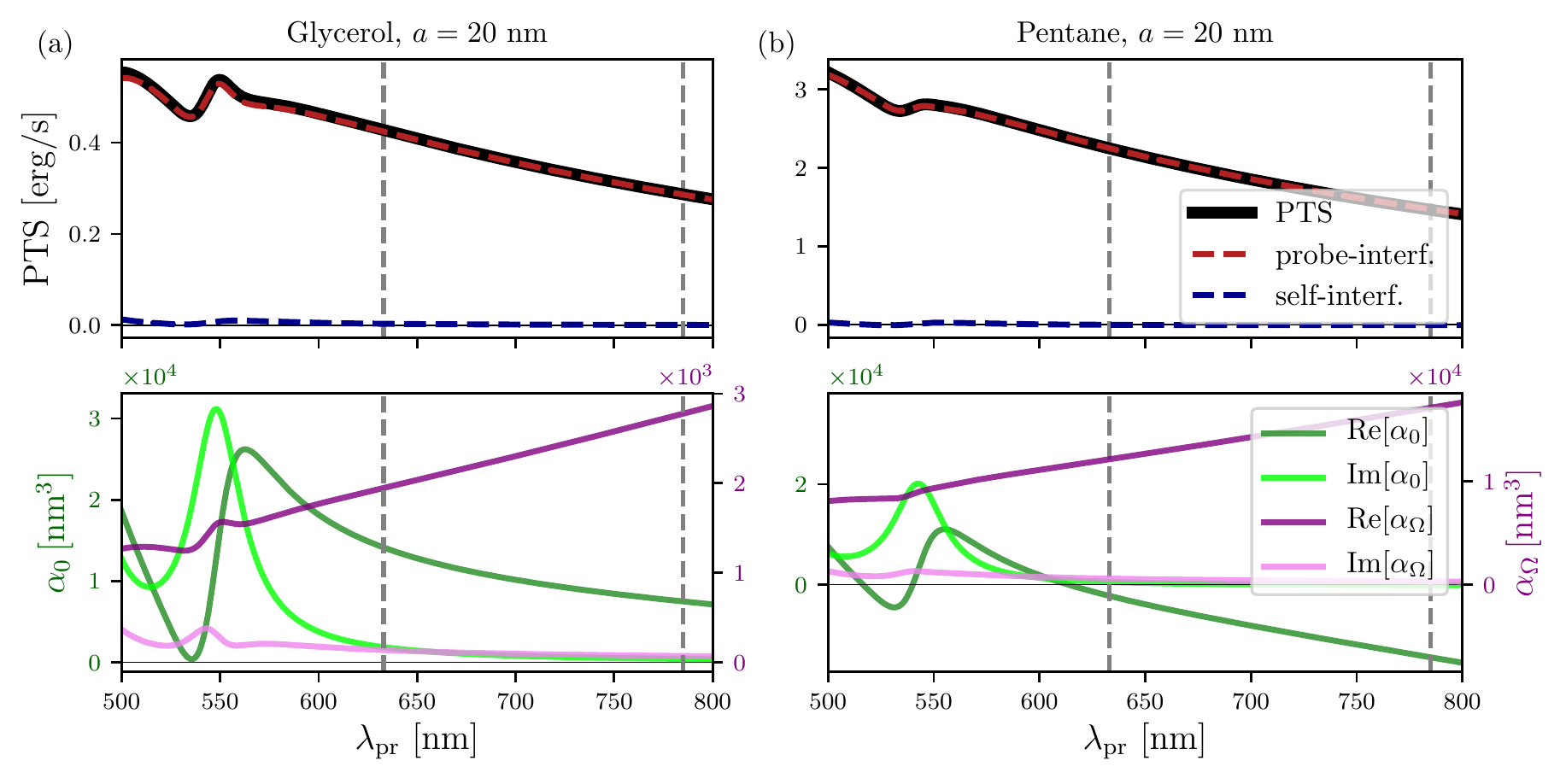} 
\caption{ \label{fig:5}
    Photothermal signal (PTS) for smaller $a=20$ nm particles shows minimal resonance effects in glycerol (a) and pentane (b), with only trivial dependence on the probe wavelength, $\propto\lambda_\mathrm{pr}^{-2}$ (see Eq.\ \eqref{eq:K_Omega}) at the two off-resonant probe wavelengths considered in Fig.\ \ref{fig:4} (dashed-gray). A small resonance effect in the PTS can be seen near the peak in $\mathrm{Im}[\alpha_0]$ (c and d, light green), but the shape of the PTS resonance is very similar to $\mathrm{Re}[\alpha_\Omega]$ (c and d, purple). The real part of the oscillating polarizability dictates the PTS as a function of $\lambda_\mathrm{pr}$, which is expected since the PTS is dominated by the probe-interference term (red) and the probe-interference is proportional to $\mathrm{Re}[\alpha_\Omega]$. Parameters used for calculation are the same as those for Fig.\ \ref{fig:4}.
    }
\end{figure}

The simple structure of the model presents a clear path to understand resonance effects, even without analyzing the complicated mathematical form of the polarizability expansion factors and the effective temperatures presented in the SI. The core-shell polarizability alters the static and modulated polarizabilities ${\alpha}_0$ and ${\alpha}_\Omega$ defined in Eqs.\ \eqref{eq:static_field} and \eqref{eq:modulated_field}. First, the metal scattering at room temperature contributes directly to the time independent polarizability $\alpha_0$ through $\alpha(T_R)$ and will increase with particle size. The static scattered field ${E}_0$ and slowly-varying scattered field ${E}_\Omega$ are therefore no longer equal, and the self-interference term will clearly grow in significance with $\alpha_0$. But even in the $a=20$ nm case discussed above where the self-interference was completely negligible, resonance effects can be seen. This effect is visible when plotting the photothermal signal as a function of probe wavelength, calculated in both glycerol (Fig.\ \ref{fig:5}a) and pentane (Fig.\ \ref{fig:5}b). Clearly the probe-interference (red) term dominates the total signal (black). A small sigmoid shape is observed in the photothermal signal near $\lambda_\mathrm{pr}\approx550$ nm, where the scattering resonance appears in Fig.\ \ref{fig:4}b. The resonance lineshape in the photothermal signal is small in the sense that its range is comparable to the variation in the photothermal signal at longer wavelengths, where Rayleigh scattering of the background dominates and $I_\mathrm{PT} \propto \lambda_\mathrm{pr}^{-2}$. Looking at the two probe wavelengths investigated above (dashed-gray), the photothermal signal has returned to Rayleigh behavior in both glycerol (Fig.\ \ref{fig:5}a) and pentane (Fig.\ \ref{fig:5}b) and shows no effect of the metal. This is consistent with the lack of $\lambda_\mathrm{pr}$ dependence in the photothermal signal trend with background plotted in Fig.\ \ref{fig:4}.

All of the dependence of the probe-interference on properties of the scatterer is contained within ${E}_\Omega$, which excludes the metal's room temperature scattering. The effect of the metal on the signal is therefore through the linear expansion coefficients ${{d}{\alpha}}/{{d}n_c}$ and ${{d}{\alpha}}/{{d}n_s}$, which will have static and time-dependent pieces, as in the background-ball case. It is intuitive that the change in polarizability with core index, ${{d}{\alpha}}/{{d}n_c}$, would be resonance dependent. Surprisingly, we find that the metal resonance also invades the shell term ${{d}{\alpha}}/{{d}n_s}$, and even dominates the signal. This is consistent with the literature precedent that the primary source of the photothermal signal is the region of heated background medium surrounding the nanoparticle, but that does not mean the metal resonance can always be neglected. Looking at the forms of shell term ${{d}{\alpha}}/{{d}n_s}$ it is clear that the resonance denominator shows up in  both the electrostatic 
(Eq.\ (S28))
and electrodynamic 
(Eq.\ (S32))
limits. This demonstrates that the in the domain of resonance effects, it may not be appropriate to interpret the photothermal signal as the superposition of signal sourced from the thermal lens and the scatter.

We can examine the role of the resonance in more detail by looking the real and the imaginary components of $\alpha_0$ and $\alpha_\Omega$ plotted in the lower half of Fig.~\ref{fig:5}. In this probe-interference dominated system, really only $\mathrm{Re}[\alpha_\Omega]$ (dark purple) impacts the signal (because $E^\mathrm{tr}_\mathrm{pr}$ is considered real). The shape of the resonance in the probe-interference and total photothermal signal in this case is therefore set by the shape of $\mathrm{Re}[\alpha_\Omega (\lambda_\mathrm{pr})]$. This polarizability does not look like, and is even an order of magnitude smaller in glycerol than the static polarizability, which looks qualitatively like the bare particle polarizability but red-shifted due to the static component of the glycerol modulation. This qualitative difference between the room temperature metal resonance and that which appears in the photothermal signal becomes much more significant as the sphere size increases. 

\begin{figure} 
\includegraphics[width=\textwidth]{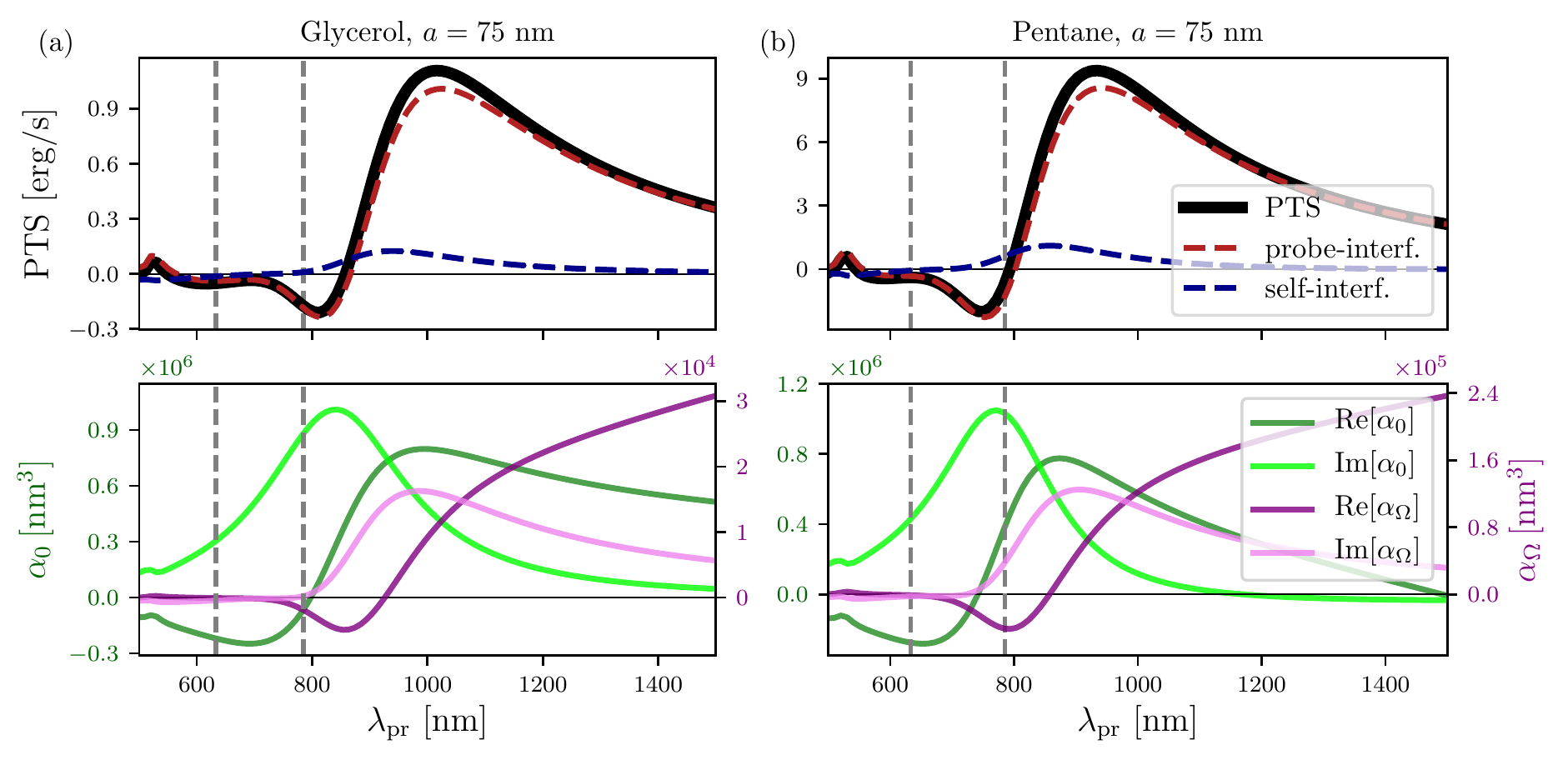}
\caption{ \label{fig:6}
    Photothermal signal (PTS) for larger $a=75$ nm particles shows a significant resonance effect with a spectral feature that blue shifts from glycerol (a) and pentane (b). Moving from glycerol to pentane, the negative peak shifts off of the $\lambda_\mathrm{pr} = 785$ nm probe line, explaining the decrease in signal seen in Fig.\ \ref{fig:4}. The self-interference term arises as a typical peak, appearing at longer wavelength than the scattering resonance. This is due to the increased magnitude of $\alpha_0 \sim 10^6$ nm$^3$ compared to the smaller particles in Fig.\ \ref{fig:5}, consistent with the room temperature polarizability increasing with particle size. The shape of the self-interference appears to follow the product of $\mathrm{Im}[\alpha_\Omega]$ (light purple, c and d) and $\mathrm{Im}[\alpha_\Omega]$, which follows from the definition of $P^\mathrm{PT}_\mathrm{SI}$. Parameters used for calculation are the same as those for Figs.\ \ref{fig:4} and \ref{fig:5}.
    }
\end{figure}

Larger particles scatter more. As one might expect, the contribution of the metal resonance to the probe-interference becomes increasingly significant as the nanoparticle size increases. Fig.\ \ref{fig:6} extends the analysis in Fig.\ \ref{fig:5} to the larger particles of radius $a=75$ nm. The sigmoidal resonance in the smaller particle photothermal signal has red-shifted and dropped below zero before returning to a broad positive peak and then falling as $\lambda_\mathrm{pr}^{-2}$. The negative peak again aligns with the typical looking peak in $\mathrm{Im}[\alpha_0]$, but clearly matches the shape of the $\mathrm{Re}[\alpha_\Omega]$, which governs the non-Rayleigh shape of the probe-interference.

It is this relatively narrow negative feature that is responsible for the $\lambda_\mathrm{pr}$ dependence of the photothermal signal trend with background visualized in Fig.\ \ref{fig:4}. Moving from glycerol to pentane, the negative peak in photothermal signal shifts off of the $\lambda_\mathrm{pr}=785$ nm line (dashed-gray), drastically decreasing the signal upon background change (seen in the difference of purple points in the top left panel of Fig.\ \ref{fig:4}). Comparing this to the $\lambda_\mathrm{pr}=633$ nm line, the signal is negligible and surprisingly flat. Resonance shifting with background will therefore have little effect in this photothermally transparent region where the blue half of the room temperature resonance occurs.

The larger particle also generates non-negligible self-interference (dashed-blue) due to the increased room temperature scattering contribution to $\alpha_0$. In considering the $y$-axis range on the lower panels of Figs.\ \ref{fig:5} and \ref{fig:6}, the ratio of $|\alpha_0|/|\alpha_\Omega|$ has increased by an order of magnitude from small to larger spheres. The self-interference appears as a typical peak, falling to zero off resonance. This self-interference resonance is redshifted from even the $\mathrm{Im}[\alpha_0]$ peak, instead coinciding with the peak in $\mathrm{Im}[\alpha_0]$, or more accurately, the product $\mathrm{Im}[\alpha_0]\cdot \mathrm{Im}[\alpha_\Omega]$. By definition of the self-interference, $P^\mathrm{PT}_\mathrm{PI} \propto k^4 (\mathrm{Re}[\alpha_0]\cdot \mathrm{Re}[\alpha_\Omega] + \mathrm{Im}[\alpha_0]\cdot \mathrm{Im}[\alpha_\Omega])$, but here the second term dominates. 

\begin{figure} 
\includegraphics[width=\textwidth]{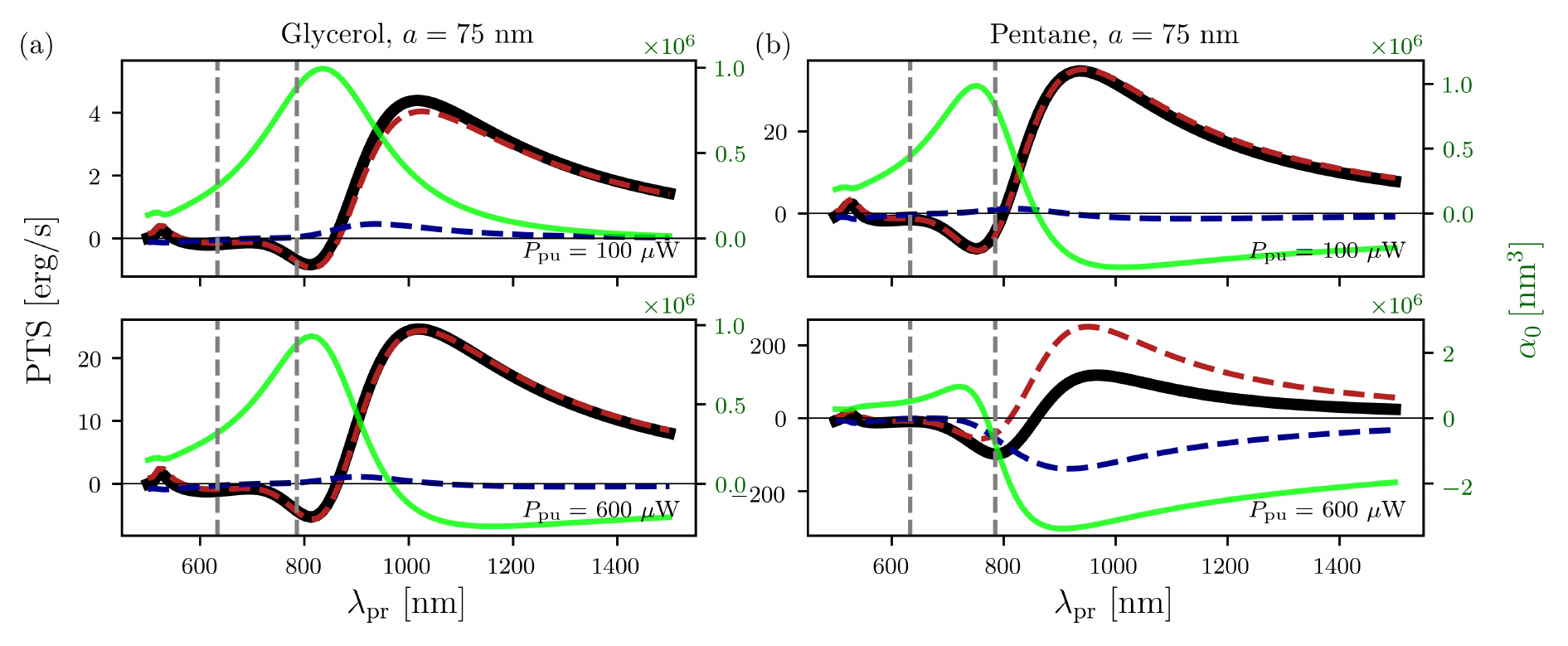} 
\caption{ \label{fig:7}
    With thermal retardation, increasing pump power does not monotonically increase self-interference as expected from the thermally static theory (Eqs.\ (\ref{eq:PI_term_L_factor}) and (\ref{eq:SI_term_L_factor})), but the effect is observed past a certain $P_\mathrm{pu}$ threshold. Upon increasing the pump power from the value of $P_\mathrm{pu} = 25$ $\mu$W used in the last several figures, the self-interference decreases to negligible values at all $\lambda_\mathrm{pr}$ by $P_\mathrm{pu} = 600$ $\mu$W in glycerol (a) and $P_\mathrm{pu} = 100$ $\mu$W in pentane (b). Continuing to increase $P_\mathrm{pu}$ does cause the self-interference to increase in magnitude, but with the opposite sign, as seen here when $P_\mathrm{pu} = 600$ $\mu$W in pentane (b). This effect is entirely governed by the the imaginary component of the static polarizability $\mathrm{Im}[\alpha_0]$ (light green, right axis), and the sign change can be traced to 
    Eq.\ (S17)
    . The room temperature scattering dominates the self-interference at low pump powers and happens to contribute positively. As the pump power increases, the heated background dominates the self-interference, which flips sign because the change in background index with temperature is negative. 
    }
\end{figure}

The metal resonance also disrupts the simple trend with pump power dependence predicted by the thermally static theory above. There it seemed that $P_\mathrm{pu}$ would modulate between significance of the probe- and self-interference contributions to the photothermal signal. The pump power dependence of the photothermal signal is explored in Fig.\ \ref{fig:7}. Increasing pump power to $P_\mathrm{pu} = 100$ $\mu$W (top row) and then $P_\mathrm{pu} = 600$ $\mu$W (bottom row), we note that the self-interference does not simply increase with $P_\mathrm{pu}$. Instead, the self-interference (dashed-blue) decreases and swings negative before finally increasing in magnitude as expected from the simple background-ball model. This effect follows $\mathrm{Im}[\alpha_0]$, which is overlaid in light green with range given by the right axis. The imaginary part of the static polarizability becomes negative with increasing $P_\mathrm{pu}$, starting with its red side, and therefore flips the sign of $P_\mathrm{SI}^\mathrm{PT}$. 

This can be explained by noting the balance between the room temperature metal scattering and the static offset contributed by the background perturbation in 
Eq.\ (S17)
. The metal perturbation term is negligible, so we need only consider the balance of the room temperature polarizability of the metal $\alpha(T_R)$ and the static offset of the background fluctuations $\propto({{d}{\alpha}}/{{d}n_s})({{d}n_s }/{{d}T}) P_\mathrm{abs} $. At low pump power, the first term will dominate, and we know that $\mathrm{Im}[\alpha(T_R)] > 0$. The background term is negligible at low pump power, where $P_\mathrm{abs}$ is small but negative because of ${{d}n_s }/{{d}T} < 0$ for all the liquid background media considered here. As the pump power, temperature, and then $P_\mathrm{abs}$ increase, the negative ${{d}{\alpha}}/{{d}n_s}$ term dominates the self-interference. After going negative, the self-interference grows in magnitude with $P_\mathrm{pu}$ as predicted in absence of the metal scattering.

The self-interference crosses zero at different pump powers for different background media. For the radius $a=75$ nm sphere in glycerol (Fig.\ \ref{fig:7}, left column), $P_{\mathrm{SI}}^{\mathrm{PT}}$ just crosses zero at $P_\mathrm{pu} = 600$ $\mu$W, but this occurs near $P_\mathrm{pu} = 100$ $\mu$W in pentane (Fig.\ \ref{fig:7}, right column). This can be attributed to the photothermal strength of pentane being significantly larger than glycerol. The background contribution to $\alpha_0$ therefore overtakes the metal contribution at lower pump power.

With all of these considerations, it is clear some of the intuition established with the simple background-ball model carries into the thermally and optically retarded regimes, as well as to larger particles with some modifications. In most parameters regimes explored here, the probe-interference dominated the total photothermal signal. The influence of the metal on the photothermal signal, within both the probe- and self-interference terms, does not affect the signal until the particle is about 20 nm in radius. At this point, the resonance appears in the photothermal signal as a function of probe wavelength as a sigmoid shape in the probe-interference. With increasing particle size and scattering cross section, the sigmoid drops the total signal to zero, eventually creating a photothermally-transparent window between approximately $\lambda_\mathrm{pr}=550-750$ nm (in glycerol). The peak emerging (here negative for chosen phase) blue of the transparent window is much narrower than the particle's scattering peak, and the signal is therefore sensitive to spectral shifts in the resonance (with consequences observed in Fig.\ \ref{fig:4} when changing background medium). It is also clear that for certain choices of the probe wavelength, the self-interference dominates the signal. This is true for a wider set of $\lambda_\mathrm{pr}$ values when the pump power has passed a certain threshold, following Eqs.\ \eqref{eq:PI_term_L_factor} and \eqref{eq:SI_term_L_factor}, and demonstrated in Fig.\ \ref{fig:7}. It is likely that in other systems, such as particles suspended in a solid \cite{selmke2012photothermal}, that the self-interference term will be more significant at smaller particle sizes. We should also emphasize that this analysis is contingent on the experimental setup modeled here (transmitted/reflected probe beam as a reference) and associated parameters chosen for calculations (discussed in figure captions). What we have seen is that the photothermal measurement depends subtly on many parameters. Therefore anyone wishing to make predictions for a specific experiment must approach the model construction and parameter definitions carefully, as well as assess the applicability of the bipolar model or any other approximations. We hope that the model presented here can form a blueprint for exploration of similar experiments.

\section{The photothermal image} \label{section:image}

The probe- and self-interference contributions to the total photothermal signal clearly have different spectral properties that will shift in relative significance with parameters such as the pump and probe wavelengths and powers. But the difference between the probe- and self-interference terms may be irrelevant to the photothermal image if the two terms are not distinguishable by their spatial form. Past literature has paid significant attention to the spatial form of the photothermal image \cite{berciaud2006photothermal,selmke2012photothermal}, as it inherits shape from both pump and probe beams. Although the simple dipole scattering model explored here will fail to capture certain aspects of the thermal lensing, we can compare the spatial form of the probe- and self-interference images to get a qualitative sense for their behavior.

In the typical experimental setup, a photothermal image is formed by rastering the nanoparticle target through the beam path. The image is therefore the photothermal signal at various positions, which we can define as the difference between the nanoparticle location $\mathbf{x}_0$ and the beam pump focal point $\mathbf{x}_f$ (assuming the pump beam is focused into the image plane, while the probe beam is focused out of plane with displacement $z_\mathrm{pr}$, as discussed in SI Section S1). The spatial dependence of the photothermal signal appears explicitly in the pump and probe intensities driving the absorption and scattering, respectively. Although both probe- and self-interference terms are linear in the probe power, but a factor of $\sqrt{P_\mathrm{pr}}$ in Eq.\ \eqref{eq:PI_term_L_factor} was contributed by the transmitted/reflected probe field, which has no spatial dependence in the raster-scan image. The probe-interference term therefore has the potentially surprising spatial dependence 
\begin{equation}
P^\mathrm{PT}_\mathrm{PI}(\mathbf{x}_f-\mathbf{x}_0) \propto \sqrt{I_\mathrm{pr}(\mathbf{x}_f-\mathbf{x}_0)}I_\mathrm{pu}(\mathbf{x}_f-\mathbf{x}_0)
\;\end{equation} 
The spatial dependence is defined by the incident beam intensities in the scattering region, which we have introduced to distinguish the local intensity from the total beam power. The spatial dependence of the self-interference term is not complicated by transmitted/reflected probe, and is therefore results trivially from Eq.\ \eqref{eq:SI_term_L_factor},
 \begin{equation}
P^\mathrm{PT}_\mathrm{SI}(\mathbf{x}_f-\mathbf{x}_0) \propto I_\mathrm{pr}(\mathbf{x}_f-\mathbf{x}_0) I_\mathrm{pu}^2(\mathbf{x}_f-\mathbf{x}_0)\;,
\end{equation}
which we note is the square of the $P^\mathrm{PT}_\mathrm{PI}$ spatial dependence. These intensities describe the focal spots of the probe and pump beams.

To qualitatively asses the differences in the contribution of $P^\mathrm{PT}_\mathrm{PI}$ and $P^\mathrm{PT}_\mathrm{SI}$ to the image, we will approximate both focal spots as that of a Gaussian beam. Taking the focal plane of both beams to contain the nanoparticle, $z = z_0 = 0$, the spatial form of each beam ($i = \textrm{pump or probe}$) then reduces from 
Eq.\ (S33)
to
\begin{equation}
I_i(r)
=
2\frac{P_i}{\pi\omega_i^2}
e^{-\frac{2r^2}{{w_i}^2}},
\end{equation}
where $r = \sqrt{x^2 + y^2}$ is the radial coordinate in the focal plane and the peak intensity has been related to the experimentally measured beam power with 
Eq.\ (S34)
. Combining this with the above expressions for the probe- and self-interference terms
\begin{align}
P^\mathrm{PT}_\mathrm{PI}(\mathbf{x}_f-\mathbf{x}_0) 
\propto{}&
\sqrt{\frac{2P_\mathrm{pr}}{\pi w_\mathrm{pr}^2}}
\frac{2P_\mathrm{pu}}{\pi w_\mathrm{pu}^2}
e^{
    -r^2
    \left(
    \frac{
        w_\mathrm{pu}^2 w_\mathrm{pr}^2
        }{
        w_\mathrm{pu}^2 + 2w_\mathrm{pr}^2
        }
        \right)^{-1}
    }
\\
P^\mathrm{PT}_\mathrm{SI}(\mathbf{x}_f-\mathbf{x}_0) 
\propto{}&
\frac{2P_\mathrm{pr}}{\pi w_\mathrm{pr}^2}
\left(\frac{2P_\mathrm{pu}}{\pi w_\mathrm{pu}^2}\right)^{\!2}
e^{
    -r^2
    \left(\frac{1}{2}
    \frac{
        w_\mathrm{pu}^2 w_\mathrm{pr}^2
        }{
        w_\mathrm{pu}^2 + 2w_\mathrm{pr}^2
        }
        \right)^{-1}
    },
\end{align} 
each term remains Gaussian, but has modified width. Approximating the resolution by the the full width at half max $\mathrm{FWHM} = 2\sqrt{2\ln{2}}\sigma$, defined by the standard deviation $\sigma$:
\begin{align}
\mathrm{FWHM}_\mathrm{PI} ={}&
2\sqrt{
\ln{2}}\frac{
    w_\mathrm{pu} w_\mathrm{pr}
    }{
    \sqrt{w_\mathrm{pu}^2 + 2w_\mathrm{pr}^2}
    }
\\
\mathrm{FWHM}_\mathrm{SI} ={}&
\frac{1}{\sqrt{2}}\mathrm{FWHM}_\mathrm{PI}.
\end{align}
The self-interference term has $\sim$ 71\% the width of the probe-interference term.

This difference in width between probe- and self-interference components of the single particle photothermal image may not be distinguishable in experiment. Especially considering that the two terms will be superimposed, this similarity between the spatial form of $P^\mathrm{PT}_\mathrm{PI}(\mathbf{x}_f-\mathbf{x}_0) $ and $P^\mathrm{PT}_\mathrm{SI}(\mathbf{x}_f-\mathbf{x}_0) $ may be the cause of the discrepancies in the published theoretical assumptions underlying the photothermal signal.

\section{Conclusions}

In this perspective, we have shown that previous assumptions behind the source of the photothermal image unify into a single theoretical model. The model contains the full complexity of the post-lock-in photothermal image, including signal contributions from interference between thermally induced scattering and the transmitted/reflected probe field (probe-interference) as well as interference between the oscillating thermal scattering and its own stationary offset (self-interference). Additional effects of target scattering resonances were also explicitly incorporated. We have shown that the general photothermal signal detected after lock-in integration reduces to common expressions used in the literature in different limits: one when neglecting the scatterer's self-interference or pure-scattering contribution, while we obtain the other common starting point by taking the thermally-static limit of our derived result. With the photothermal signal being the sum of probe- and self-interference components, we demonstrated how to increase the model's complexity to increase accuracy while tracing intuition from the simplest form of the model. This approach revealed that probe- and self-interference play different roles in the photothermal spectrum, particularly in probe wavelength. However, with so many parameter dependencies, isolating the role of these two terms in a given experiment may be difficult. Differentiating probe- and self-interference may not be much simpler in terms of the spatial dependence of the photothermal image, with the two terms differing only slightly in width. This minimal difference in the two images paired with the complexity of the parameter space that tunes the relative contributions of probe- and self-interference may explain the past success of using either term in modeling different experiments. Nevertheless, with the knowledge gained from the presented model we expect the resonance effects and photothermally-transparent spectral windows of scattering particles to be exploitable for further optimization. Taken together, the understanding presented herein provides a stepping stone to quantitative correlation of the measured photothermal signal and the local temperature, leading towards an all-optical nano-thermometer. This model additionally lays the foundation for understanding the photothermal images of plasmonic aggregates, with the aim of spatially resolving their temperature profiles in experiment and guiding the design of future thermal metamaterials.

\section{Supplementary Material}
See the supplementary material for (S1) a model of the probe interference contribution to the photothermal signal; (S2) the effects of thermal retardation in both the core and shell; (S3) the temperature dependence of the polarizability; (S4) the functional form of the Gaussian probe beam; (S5) the modified long wavelength approximation polarizability; the incorporation of (S6) temperature-dependent refractive index and (S7) substrate effects upon the polarizability; (S8) an alternative ansatz for the modulated pump beam; and (S9) the dependence of the photothermal signal upon the particle–probe-focus offset at different probe wavelengths and in different solvents.


\section{Acknowledgments}
This work was supported by the U.S. National Science Foundation under grant nos. NSF CHE-1727092 (D.J.M.) and CHE-1727122 (S.L.). S.L. also acknowledges support from the Robert A. Welch Foundation (grant no. C-1664).

\bibliography{refs}

\end{document}


\renewcommand{\thesection}{S\arabic{section}}
\renewcommand{\theequation}{S\arabic{equation}}
\renewcommand{\thefigure}{{S\arabic{figure}}}

\title{Supplementary information: Resolving resonance effects in the theory of single particle photothermal imaging}

\author{Harrison J. Goldwyn}
\email{goldwyn@uw.edu}
\affiliation{Department of Chemistry, University of Washington, Seattle, WA 98195}
\author{Stephan Link}
\affiliation{Department of Chemistry, Rice University, Houston, TX 77005}
\affiliation{Department of Electrical and Computer Engineering, Rice University, Houston, TX 77005}
\author{David J. Masiello}
\email{masiello@uw.edu}
\affiliation{Department of Chemistry, University of Washington, Seattle, WA 98195}


\maketitle

\section{Model of probe interference contribution} \label{section:transmitted_probe_field}

Neglecting differences in the spatial modulation of the probe and scattered fields, the transmitted/reflected probe field in the collection region can be approximated to have the spatial form of a radiating dipole with effective polarizability or scattering volume $V$. Using the scalar notation presented in the main text to omit the spatial dependence of the fields that is integrated out at the detector, 
\begin{equation}
{E}^\mathrm{tr}_\mathrm{pr}
\approx
{G} \cdot V \cdot {E}_\mathrm{pr}(\mathbf{x}_0).
\end{equation}
The product of effective scattering volume and the probe field in the focal spot can be parameterized by the experimentally measured probe power. Defining the collection aperture in the background medium by the polar angle $\theta_{\mathrm{col}} = \sin^{-1}(\mathrm{NA}_{\mathrm{col}}/n_b)$, the power radiated by an electric dipole oriented perpendicular to the optical axis through the solid angle $(\theta_{\mathrm{col}}, 2\pi)$ is
\begin{align}
P_\mathrm{scatt}(\theta_{\mathrm{col}}) 
={}&
\int_0^{\theta_{\mathrm{col}}} \int_0^{2\pi} r^2 \sin\theta\,{d}\theta  {d}\phi \, I_\mathrm{scatt}(r, \theta, \phi) 
\\
={}&  \int_0^{\theta_{\mathrm{col}}} {d}\theta r^2 \sin\theta \int_0^{2\pi}  {d}\phi \,\frac{c k_\mathrm{pr}^4}{3}|\mathbf{p}|^2 \frac{1 - \sin^2\theta \cos^2\phi}{r^2}
\\
={}& \frac{c k_\mathrm{pr}^4}{3}|\mathbf{p}|^2 f(\theta_{\mathrm{col}}).
\end{align} 
The $\theta_{\mathrm{col}}$-dependent factor that determines the fraction of total scattered dipole radiation that makes it through the aperture is
\begin{equation} \label{eq:power_scatt_collection_angle}
f(\theta) = 
\left[
    \frac{3}{4}(1-\cos\theta)
    - 
    \frac{1}{2}(2+\cos\theta)\sin^4\frac{\theta}{2}
    \right]
\;,\end{equation} 
which increases monotonically from 0 to 1/2 for $0\leq\theta\leq\pi/2$. For a background medium of glycerol and a numerical aperture of $\mathrm{NA}_\mathrm{col} = 0.7$, $f(\theta_{\mathrm{col}}) = 0.085$.

Equating the scattered power from the effective dipole $|p| = V \cdot |{E}_\mathrm{probe}(\mathbf{x}_0)|$ to the experimentally measured probe power yields,
\begin{equation}
V \cdot |{E}_\mathrm{probe}(\mathbf{x}_0)| 
\approx
\sqrt{
    \frac{1}{f(\theta_{\mathrm{col}})} \frac{3 P^\mathrm{exp}_\mathrm{pr}}{c k_\mathrm{pr}^4}
    }
\;.\end{equation}
The transmitted probe field magnitude at the detector then takes the form
\begin{align} \label{eq:trans_probe_field}
|{E}^\mathrm{tr}_\mathrm{pr}|
\approx{}&
{G} \cdot \sqrt{
    \frac{1}{f(\theta_{\mathrm{col}})} \frac{3 P^\mathrm{exp}_\mathrm{pr}}{c k_\mathrm{pr}^4}
    }
=
k_\mathrm{pr}
\sqrt{
    \frac{1}{f(\theta_{\mathrm{col}})} \frac{3 P^\mathrm{exp}_\mathrm{pr}}{c}
    }
\;,\end{align}
where in the last line we have used the fact that in the scalar notation $G = k_\mathrm{pr}^3$. The $k_\mathrm{pr}$ (and therefore $n_b$) dependence of ${E}^\mathrm{tr}_\mathrm{pr}$ does not propagate to the observable, as it will cancel with denominator in the integrated expression $P_\mathrm{PT}^\mathrm{raw}(t) = I_\mathrm{PT}^\mathrm{raw} f(\theta_\mathrm{col}) / k_\mathrm{pr}^2$.

Past modeling of the photothermal image has demonstrated the significance of the difference in phase between the reference beam and the scattered field \cite{li2020resonant}, which leads to elimination of the photothermal signal when the probe focus nears the nanoparticle \cite{selmke2012photothermal}. To reproduce this effect in our model, we account for the Gouy phase shift of $\pi/2$ inherited by the expanding probe field between the nanoparticle and the detector \cite{HWANG2007487}. Finally, the transmitted probe field at the detector takes the following form in our integrated shorthand, 
\begin{equation}
{E}^\mathrm{tr}_\mathrm{pr} \approx |{E}^\mathrm{tr}_\mathrm{pr}| e^{-i(kz_\mathrm{pr} - \frac{\pi}{2})}
\end{equation} 
where $z_\mathrm{pr}$ is the distance from the particle to the probe beam focus \cite{selmke2015physics}.

\section{Accounting for thermal retardation in the core and shell temperature}
\label{section:thermal_retardation_details}

For a sphere of radius $a$ in an isotropic background medium absorbing at a rate, 
\begin{equation}
P_\mathrm{abs}(t) = P_\mathrm{abs} \frac{1+\cos{\Omega{t}}}{2}
\end{equation}
the temperature increase from room temperature is determined by solving the time-dependent heat diffusion equation outside the sphere with the boundary condition at its surface,
\begin{equation} \label{eq:time_dependent_sphere_boundary_condition}
P_\mathrm{abs} \frac{1+\cos{\Omega{t}}}{2} 
=
- 4 \pi \kappa a^2 
\left.
\frac{\partial{T}}{\partial{r}} 
\right|_{r=a}
\;,
\end{equation} 
which equates the heat flow out of the sphere following absorption to the integral of the temperature gradient across the sphere surface. This boundary condition assumes a uniform temperature on the sphere, which is accurate for noble metal nanoparticles with thermal constants much greater than those of the background and smaller in size than the thermal radius. The solution satisfying this boundary condition as well as the radial heat diffusion equation ($r=|{\bf x}|$),
\begin{equation}
r^2 \frac{\partial}{\partial{t}}T
-
\frac{\kappa}{C_p}
\frac{\partial}{\partial{r}}
\left(
    r^2 \frac{\partial}{\partial{r}} T
    \right)
= 
0
\end{equation}
can be written as
\begin{equation} \label{eq:time_dep_heat_sln}
\Delta T(r, t) ={}
\frac{P_\mathrm{abs}}{4\pi\kappa{r}}
\cdot \frac{1}{2}\left(
    1
    +
    \frac{
        e^{-\frac{r-a}{r_\mathrm{th}}}
        }
        {
        (\frac{r_\mathrm{th} + a}{r_\mathrm{th}})^2 + (\frac{a}{r_\mathrm{th}})^2
        }
        \left[
            \frac{r_\mathrm{th} + a}{r_\mathrm{th}} \cos(\Omega{t} - \frac{r-a}{r_\mathrm{th}})
            +
            \frac{a}{r_\mathrm{th}} \sin(\Omega{t} - \frac{r-a}{r_\mathrm{th}})
            \right]
    \right),
\end{equation}
where the temperature inside the sphere is defined by $r \to a$ for $r < a$. This spatially varying expression is coarsely approximated by a shell of uniform temperature. To assign an effective temperature, we volume average the temperature over the shell of radius $b$,
\begin{align}
\langle \Delta T(r, t) \rangle_{V_\mathrm{shell}} 
={}& 
\frac{4\pi}{V_\mathrm{shell}}
\int_r\,r^2\Delta T(r, t)
\\
={}& \nonumber
\frac{P_\mathrm{abs}}{4 \kappa V_\mathrm{shell}}
\left[
    b^2 - a^2
    +
    r_\mathrm{th}^2 \sin(\Omega{t})
    \right.
    \\ \nonumber
    &\phantom{
        \frac{P_\mathrm{abs}}{4 \kappa V_\mathrm{shell}}[]
        }
    -
    \frac{
        e^{-\frac{b-a}{r_\mathrm{th}}}
        }
        {
        (\frac{r_\mathrm{th} + a}{r_\mathrm{th}})^2 + (\frac{a}{r_\mathrm{th}})^2
        }
    \left(
        r_\mathrm{th}(b-a)
        \cos(\Omega{t} - \frac{b-a}{r_\mathrm{th}})
        \right.
        \\
        &\phantom{
            \frac{P_\mathrm{abs}}{4 \kappa V_\mathrm{shell}}[]
            \frac{
                    e^{-\frac{b-a}{r_\mathrm{th}}}
                    }
                    {
                    (\frac{r_\mathrm{th} + a}{r_\mathrm{th}})^2 + (\frac{a}{r_\mathrm{th}})^2
                    }
            }
        \left.
        \left.
        + 
        (r_\mathrm{th}(b+r_\mathrm{th}) + a(2b+r_\mathrm{th})  )
        \sin(\Omega{t} - \frac{b-a}{r_\mathrm{th}})
        \right)
    \right].
\end{align}
Defining this value as the shell temperature $\Delta T_\mathrm{shell} = \langle \Delta T(r, t) \rangle_{V_\mathrm{shell}}$ and the core temp to be 
\begin{equation}
\Delta{T}_\mathrm{core} = \Delta T(r=a, t)
=
\frac{P_\mathrm{abs}}{4\pi\kappa{a}}
\cdot \frac{1}{2}\left(
    1
    +
    \frac{
        \frac{r_\mathrm{th} + a}{r_\mathrm{th}} \cos(\Omega{t})
            +
            \frac{a}{r_\mathrm{th}} \sin(\Omega{t})
        }{(\frac{r_\mathrm{th} + a}{r_\mathrm{th}})^2 + (\frac{a}{r_\mathrm{th}})^2}
    \right)
\end{equation}
the core-shell polarizability (to first order in $T$) can be written generally to first order in temperature fluctuations as 
\begin{equation} \label{eq:alpha_exp}
\alpha_\mathrm{cs} (T)
=
\alpha(T_R)
+
\frac{{d}{\alpha}}{{d}n_c }
\frac{{d}n_c }{{d}T} 
        \Delta T_\mathrm{core}
+
\frac{{d}{\alpha}}{{d}n_s}
\frac{{d}n_s }{{d}T} 
    \langle \Delta T(r, t) \rangle_{V_\mathrm{shell}} 
\;,
\end{equation}
which we have found agrees well with experimental data. Separating the polarizability into static and time varying components, $\alpha_\mathrm{cs}(t) = \alpha_0 + {\alpha}_\Omega(t)$, and noting that $V_\mathrm{shell} = \frac{4\pi}{3}(b^3-a^3)$, the time-independent piece can be written as
\begin{align}
\alpha_0 = \label{eq:alpha_0}
\alpha(T_R)
+ 
    \frac{{d}{\alpha}}{{d}n_c} \frac{{d}n_c }{{d}T}
    \frac{P_\mathrm{abs}}{8\pi\kappa{a}} 
+
  \frac{{d}{\alpha}}{{d}n_s}
    \frac{{d}n_s }{{d}T} 
    \frac{3 P_\mathrm{abs}}{16 \pi \kappa}
    \frac{b^2 - a^2}{b^3 - a^3}
\end{align}
and the time-dependent (fluctuational) polarizability is 
\begin{align}
{\alpha}_\Omega(t) ={}& \nonumber
    \frac{{d}{\alpha}}{{d}n_c} \frac{{d}n_c }{{d}T} 
        \frac{P_\mathrm{abs}}{8\pi\kappa{a}}
        \frac{
            \frac{r_\mathrm{th} + a}{r_\mathrm{th}} \cos(\Omega{t})
                +
                \frac{a}{r_\mathrm{th}} \sin(\Omega{t})
            }{(\frac{r_\mathrm{th} + a}{r_\mathrm{th}})^2 + (\frac{a}{r_\mathrm{th}})^2}
\\ \nonumber
&+
  \frac{{d}{\alpha}}{{d}n_s}
    \frac{{d}n_s }{{d}T} 
    \frac{3 P_\mathrm{abs}}{16 \pi \kappa (b^3 - a^3)}
    \\ \nonumber
    &\phantom{+}\,
    \times
    \left[
    r_\mathrm{th}^2 \sin(\Omega{t})
    -
    \frac{
        e^{-\frac{b-a}{r_\mathrm{th}}}
        }
        {
        (\frac{r_\mathrm{th} + a}{r_\mathrm{th}})^2 + (\frac{a}{r_\mathrm{th}})^2
        }
    \left(
        r_\mathrm{th}(b-a)
        \cos(\Omega{t} - \frac{b-a}{r_\mathrm{th}})
        \right.
        \right.
        \\
        &\phantom{
            \frac{P_\mathrm{abs}}{4 \kappa V_\mathrm{shell}}[]
            \frac{
                    e^{-\frac{b-a}{r_\mathrm{th}}}
                    }
                    {
                    (\frac{r_\mathrm{th} + a}{r_\mathrm{th}})^2 + (\frac{a}{r_\mathrm{th}})^2
                    }
            }
        \left.
        \left.
        + 
        (r_\mathrm{th}(b+r_\mathrm{th}) + a(2b+r_\mathrm{th})  )
        \sin(\Omega{t} - \frac{b-a}{r_\mathrm{th}})
        \right)
    \right]
\\ \nonumber
\equiv{}& 
{\alpha}_\Omega^{(0)} \cos(\Omega{t})
+
{\alpha}_\Omega^{(\pi/2)} \sin(\Omega{t})
\\ 
&+
{\alpha}_\Omega^{(\phi_{r_\mathrm{th}})}\cos(\Omega{t} - \frac{b-a}{r_\mathrm{th}})
+
{\alpha}_\Omega^{(\pi/2 + \phi_{r_\mathrm{th}})} \sin(\Omega{t} - \frac{b-a}{r_\mathrm{th}})
\end{align}
where we have defined four components to the polarizability based on their phase delay from the absorption process. 

We can now express the scattered field in terms of its static and fluctuating components just as in the thermally-static case,
\begin{align}
\mathbf{E}_\text{scattered}(t) 
&= \mathbf{G} \cdot \left(
    \boldsymbol{\alpha}_0  
    + 
    {\boldsymbol{\alpha}_\Omega}(t)
    \right)
\cdot\mathbf{E}_\mathrm{pr}(\mathbf{x}_0) e^{-i \omega t}
\\ \nonumber
&\equiv{}
\left[ \mathbf{E}_0  +  \mathbf{E}_\Omega (t) \right] e^{-i \omega t}
\;,
\end{align}
which we can then use to calculate the lock-in-integrated photothermal signal as above, this time with a generalized phase delay $\phi$, 
\begin{align}
P^\text{PT} 
\equiv{}&
\lim_{\tau\to\infty} \frac{1}{\tau}\int_0^\tau  {d}t\,  I_\mathrm{PT}^\mathrm{raw} (t) \cos(\Omega t - \phi)
\\
={}& 
    \frac{c n}{8\pi}
    \lim_{\tau\to\infty} 
    \frac{1}{\tau}\int_0^\tau {d}t\, \left|
        \left[ 
            \mathbf{E}^\mathrm{tr}_\mathrm{pr} +  \mathbf{E}_0  +  \mathbf{E}_\Omega (t)
            \right]
        \right|^2
    \cos(\Omega t - \phi)
    \\
={}& \nonumber
    \frac{c n}{8\pi}
    \mathrm{Re}\Big[ (\mathbf{E}^\mathrm{tr}_\mathrm{pr} + \mathbf{E}_0)^* \cdot
        \Big(
        \mathbf{E}_\Omega^{(0)}\cos(\phi)
        +
        \mathbf{E}_\Omega^{(\pi/2)}\sin(\phi)
\\ &\phantom{
    \frac{c n}{8\pi}\mathrm{Re}\Big[ (\mathbf{E}^\mathrm{tr}_\mathrm{pr} + \mathbf{E}_0)^* \cdot\Big(}
    +
    \mathbf{E}_\Omega^{(\phi_{r_\mathrm{th}})}\cos(\phi - \frac{b-a}{r_\mathrm{th}})
    +
    \mathbf{E}_\Omega^{(\pi/2 + \phi_{r_\mathrm{th}})}\sin(\phi - \frac{b-a}{r_\mathrm{th}})
        \Big)
        \Big]
\\
={}&
\frac{c n}{8\pi}
    \mathrm{Re}\Big[ 
        (\mathbf{E}^\mathrm{tr}_\mathrm{pr} + \mathbf{E}_0)^* 
        \cdot
        \mathbf{E}_\Omega(t=\phi/\Omega)
    \Big]
\end{align}
where the four field components are defined by $\mathbf{E}_\Omega^{i} = \mathbf{G}\cdot \boldsymbol{\alpha}_\Omega^i \cdot \mathbf{E}_\mathrm{pr}(\mathbf{x}_0)$.

\section{Temperature dependence of the polarizability}\label{section:alpha_coreshell}

Here we detail the core-shell polarizability used to model the scattering by the nanoparticle and its surrounding region of heated background.  
We start with a quasistatic model appropriate for particles much smaller than the relevant wavelength of light and then introduce a retardation-corrected model, in what is often called the modified long wavelength approximation (MLWA) \cite{meier1983enhanced,moroz2009depolarization}.
In either case, the heated background surrounding each particle is approximated by a spherical shell around the spherical metal core. 
The metal is assigned a temperature-dependent dielectric function by fitting a Drude-Lorentz model to ellipsometry data \cite{zahra} (discussed further in SI Section {\bf II}). All background media are assumed to have optical properties that vary linearly with temperature, with values specified in SI Table 1.

In the quasistatic limit, the polarizability of these core-shell model particles can be written in closed form, as derived electrostatically in reference \cite{bohren2008absorption},
\begin{equation} \label{eq:alpha_cs_QS}
\alpha_{cs} = b^3\frac
    {
    (\epsilon_s-\epsilon_b)(\epsilon_c+2\epsilon_s)+f(\epsilon_c-\epsilon_s)(\epsilon_b+2\epsilon_s)    
    }
    {
    (\epsilon_s+2\epsilon_b)(\epsilon_c+2\epsilon_s)+f(2\epsilon_s-2\epsilon_b)(\epsilon_c-\epsilon_s)
    },
\end{equation}
where $\epsilon_s$ is the electric permeability of the heated background shell, $\epsilon_c$ is the permeability of the metal core, and $f=a^3/ b^3$ is the ratio of core volume to total core-shell volume.

To model the temperature dependence of the scattering process, the polarizability will be expanded to linear changes in temperature. 
We have found that to reasonable accuracy the refractive indices for both the metal core (with complex index $n_c = n_c^\prime + i n_c^{\prime\prime} = \sqrt{\epsilon_c}$) and the background shell (real index $n_s = \sqrt{\epsilon_s}$) can be treated as linear in temperature across the range of temperatures reachable in relevant experiments.
The rate of change with temperature is approximately ${{d}n}/{{d}T}\approx 10^{-4}$ for all background media considered here \cite{gaiduk2010detection}, as well as the metal. 
This small rate of refractive index change facilitates the expansion of the polarizability $\alpha_{cs}$ to first order in change of all refractive indices with temperature. 
With $\Delta{n} \equiv{{d} n}/{{d}T}|_{T=T_R}$ small for all $n =\{ n_b, n_c^\prime, n_c^{\prime\prime}\}$, the expanded polarizability written generally in Eq.\ \eqref{eq:alpha_exp} is justified. 

In the electrostatic limit, the core-shell polarizability in Eq.\ \eqref{eq:alpha_cs_QS} expanded to linear order yields the coefficients 
\begin{align}
\alpha_\mathrm{cs}^\mathrm{QS} (T_R) 
&= \label{eq:alpha_cs^QS}
a^3 \frac{\epsilon_c -\epsilon_b}{\epsilon_c + 2\epsilon_b} 
\\
\frac{{d}{\alpha^\mathrm{QS}}}{{d}n_c}
&=\label{eq:aalphadnc_cs^QS}
6 a^3 \frac{ \epsilon_b}{(\epsilon_c + 2\epsilon_b)^2} n_c
\\
\frac{{d}{\alpha^\mathrm{QS}}}{{d}n_s}
&=\label{eq:aalphadns_cs^QS}
\frac{2 (b^3-a^3) }{3} 
    \left[1 + 2 a^3/b^3 \left(\frac{\epsilon_c -\epsilon_b}{\epsilon_c + 2\epsilon_b} \right)^{\!\!2} \right]
    \frac{1}{n_s}
.\end{align}
It is clear in that the room temperature contribution is simply the Clausius-Mossotti relation, well known to give the static polarization response of a sphere. The core contribution notably scales with the volume of the core similar to the room temperature polarizability and has a denominator with similar poles. The shell contribution scales overall with the volume of the shell, and contains a linear combination of a core-independent piece with a term that depends on the core resonance and scales with the core volume.

In the electrodynamic limit, the core-shell polarizability can be corrected for retardation effects to the resonance structure in the MLWA. The procedure for solid spherical and spheroidal nanoparticles is well documented \cite{meier1983enhanced,moroz2009depolarization}, and provides an expression for the retardation corrected polarizability as a function of the static polarizability and the wave vector magnitude $k=\omega n/c$. This procedure was extended in 2009 to shelled particles \cite{chung2009dynamic} by finding an effective permittivity to fill a solid particle such that this hypothetical particle has equivalent static polarizability to the shelled particle. With this new effective polarizability, the solid-particle formula for the MLWA polarizability can be used. In our case, for the sphere with core radius $a$ and shell radius $b$, the MLWA polarizability can be shown to be
\begin{equation} \label{eq:alpha_cs_MLWA}
\alpha_\mathrm{cs}^\mathrm{ML} 
= 
\frac{b^3}{3} 
\frac{
    (\epsilon_s - \epsilon_b) (\epsilon_c q_c - \epsilon_s [q_c - 1]) b^3
     -
     (\epsilon_c - \epsilon_s) (\epsilon_s [q_s - 1] - \epsilon_b q_s) a^3
    }{
    (\epsilon_c q_c - \epsilon_s [q_c - 1]) (\epsilon_s q_s - \epsilon_b [q_s - 1]) b^3
     -
     (\epsilon_c - \epsilon_s) (\epsilon_s - \epsilon_b) q_s (q_s - 1) a^3
    }
\;,
\end{equation}
where the retardation factors $q_i = \frac{1}{3}[1 - (ka_i)^2- i \frac{2}{3} (ka_i)^3]$ are defined by the core and shell radii, $a_c = a$ and $a_s = b$. The coefficients resulting from a linear expansion of Eq.\ \ref{eq:alpha_cs_MLWA} are as follows, employing the simplifying notation $\epsilon \equiv \epsilon_c/\epsilon_b$,
\begin{align}
\alpha^\mathrm{ML} (T_R) 
={}& \label{eq:alpha_cs^ML} 
\frac{a^3}{3} \frac{\epsilon - 1}{ q_c [\epsilon + 1] - 1} 
\\
\frac{{d}{\alpha^\mathrm{ML}}}{{d}n_c}
={}&\label{eq:aalphadnc_cs^ML}
\frac{2}{3} a^3 \frac{ 1 }{\epsilon q_c + [q_c - 1]^2}  n_c
\\
\frac{{d}{\alpha^\mathrm{ML}}}{{d}n_s}
={}&\label{eq:aalphadns_cs^ML}
\frac{
    n_b
    }{
    3 b^3 ( q_c [1 - \epsilon] - 1 )     
    }
    \Big[
    b^6 
    (q_c[1 - \epsilon] - 1)^2 
    + 
    a^6 
    (\epsilon - 1)^2 (q_s-1) q_s
    \\\nonumber
    &{}+ 
    a^3b^3
    (
        \epsilon^2 q_c [1-2q_s]
        +
        [q_c -1 + 2q_s - 2q_c q_s]
        + 
        2 \epsilon [ -q_s + q_c ( 2 q_s - 1) ]
        )
    \Big]
.\end{align}
Here, the room temperature contribution and core temperature contribution both follow from above. The room temperature contribution is the standard MLWA polarizability for a sphere, and the core fluctuation coefficient takes the same relationship to the room temperature polarizability as it does in the quasistatic limit. The form of the shell fluctuation term is difficult to simplify, but may reveal similar structure to its quasistatic analog.

Even for small particles ($a\sim 10$ nm), we found the MLWA polarizability to yield a qualitatively different photothermal signal (as a function of probe wavelength) than the quasistatic polarizability. This is likely due to the fact that, in calculations, we have set the radius of the background shell $b$ equal to the probe beam waist $w_\mathrm{pr}$ to represent the optically active region of heated background medium in the scattering part of the problem. This radius is on the order of 100's of nm on resonance with the metal, which is likely big enough to break the quasistatic assumptions. All of the calculations presented in this Perspective therefore use the MLWA model and the quasistatic polarizability is presented only for completeness and analogy.

\section{Gaussian beam}\label{section:gau_beam}

For simplicity, we assume the focused probe field in the sample region takes the form of a Gaussian beam \cite{svelto2010principles}. This approximation is known to fail for large numerical aperture objectives, but we proceed with the Gaussian beam as a qualitative predictor of effects related to focal spot size. The electric field in the focal spot is defined by, 
\begin{equation}
\mathbf{E}(r,z)
= \label{eq:gau_beam}
E_0 \hat{\bf e}_x \frac{w_0}{w(z)}
e^{-\frac{r^2}{{w(z)}^2}}
e^{-i\left(kz + k\frac{r^2}{2{R(z)}^2} - \psi(z)\right)}
\;,
\end{equation}
where $w(z) = w_0 \sqrt{1 + (\frac{z}{z_R})^2}$ is the beam waist along the optical axis ($z$). $z_R = {\pi w_0^2 n}/{\lambda}$ defines the Rayleigh length in terms of the focused beam waist $w_0$, the background refractive index $n$ and the vacuum wavelength $\lambda$, which marks the distance from the focal plane where the beam waist falls to $\sqrt{2}w_0$ and the intensity is half its peak values. The wavefront radius of curvature is defined by $R(z) = z[1 + ({z}/{z_R})^2]$ and the Gouy phase is $\psi = \arctan({z}/{z_R})$. 

The field magnitude is defined by the peak intensity at the center of the focal spot $I_0 =({cn}/{8\pi})|E_0|^2$. This peak intensity is simply related to the total power transmitted by 
\begin{align} \label{eq:E0_gaussian_defbyPexp}
\frac{I_0}{2} ={}& \frac{P_\mathrm{exp}}{\pi w_0^2}
\;,
\end{align}
which we can use to define the field amplitude in terms of an experimentally determined laser power,
\begin{align}
E_0 ={}& \frac{4}{w_0} \sqrt{\frac{P_\mathrm{exp}}{ {cn} }}
\;.
\end{align}

For all calculations presented, the pump and probe beam waists are assigned by the numerical aperture of the illumination objective by equating the FWHM to the an experimentally realizable resolution $\mathrm{FWHM} = 0.61({\lambda_i}/{\mathrm{NA}_\mathrm{illu}}) = \sqrt{2\ln{2}}w_i$, 
\begin{equation}
w_i = \frac{0.61}{\sqrt{2\ln{2}}}\frac{\lambda_i}{\mathrm{NA}_\mathrm{illu}}
=
0.52
\frac{\lambda_i}{\mathrm{NA}_\mathrm{illu}}\;.
\end{equation}

\begin{figure} 
\includegraphics[width=\textwidth]{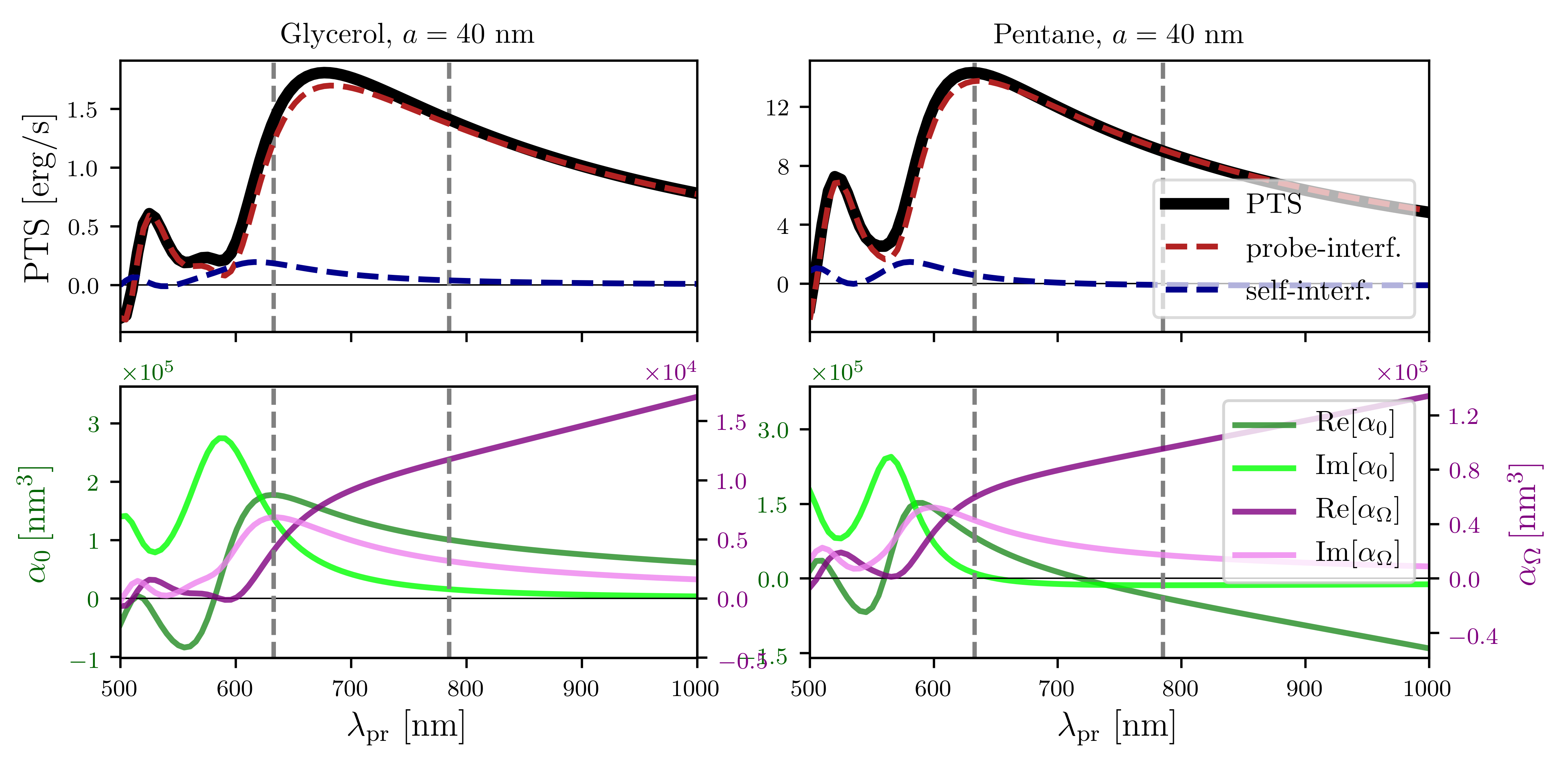} 
\caption{ \label{fig:4_a40}
    Photothermal signal (top) for $a=40$ nm particles in glycerol (left) and pentane (right) shows the evolution of the photothermal resonance with increasing particle size when compared to Figs.\ 4 and 5 in the main text (for $a=20$ and $a=75$ nm particles respectively). Similar analysis to that in the caption of Fig.\ 5 can be applied here to connect the components of polarizability with the probe- and self-interference terms. 
}
\end{figure}

\section{Absorbing polarizability} \label{section:abs_appendix}

For all calculations, the absorbing polarizability was assigned the exact dipolar Mie polarizability, 
\begin{align}
\alpha_\mathrm{Mie} 
={}& 
\frac{i 3}{2k^3} a_\mathrm{Mie}  
\end{align}
where the dipole Mie coefficient is 
\begin{equation}
a_\mathrm{Mie} =\frac{
    \epsilon_r^2 j_1(\epsilon_rka)(j_1(ka) +ka j_1^\prime(ka)) 
    - 
    j_1(ka) j_1(\epsilon_r ka) + \epsilon_r ka  j_1^\prime(\epsilon_r ka)
    }{
    \epsilon_r^2 j_1(\epsilon_rx)(h_1(ka) + ka h_1^\prime(ka)) 
    - 
    h_1 ka (j_1(\epsilon_r ka) + \epsilon_r ka  j_1^\prime(\epsilon_r ka))
    }
\end{equation}
in terms of the relative permittivity $\epsilon_r = \epsilon_c/\epsilon_b$, the wave vector magnitude $k=\omega_\mathrm{pu} n_b / c$ and the spherical Bessel and Hankel functions $j$ and $h$. To better see the dependence of the background optical properties, we can use the MLWA polarizability with little loss in accuracy \cite{meier1983enhanced},
\begin{align}
\alpha^\mathrm{ML}_\mathrm{sphere} 
= 
\frac{a^3}{3} 
\frac{
    \epsilon_c(\omega_\mathrm{pu}) - \epsilon_b
    }{
    \epsilon_c(\omega_\mathrm{pu}) q_c(\omega_\mathrm{pu}) + \epsilon_b[q_c(\omega_\mathrm{pu}) - 1]
    }
\end{align}
where $q_c = \frac{1}{3}[1 - (ka)^2- i \frac{2}{3} (ka)^3]$.

\section{Change in metal refractive index with temperature}
\label{section:metal_dndT}

The metal dielectric function was modeled by fitting a Drude-Lorentz model to temperature dependent dielectric data of gold from Bilchak and Fakhraai \cite{zahra}. The data consisted of  ellipsometry determined vales for the real and imaginary parts of the refractive index at 1 K increments from room temp to +100 K at wavelengths between 550 and 1047 nm. A separate model was fit at each temperature, emphasizing accuracy at small wavelengths near the small particle resonance. 

\begin{figure} 
\includegraphics{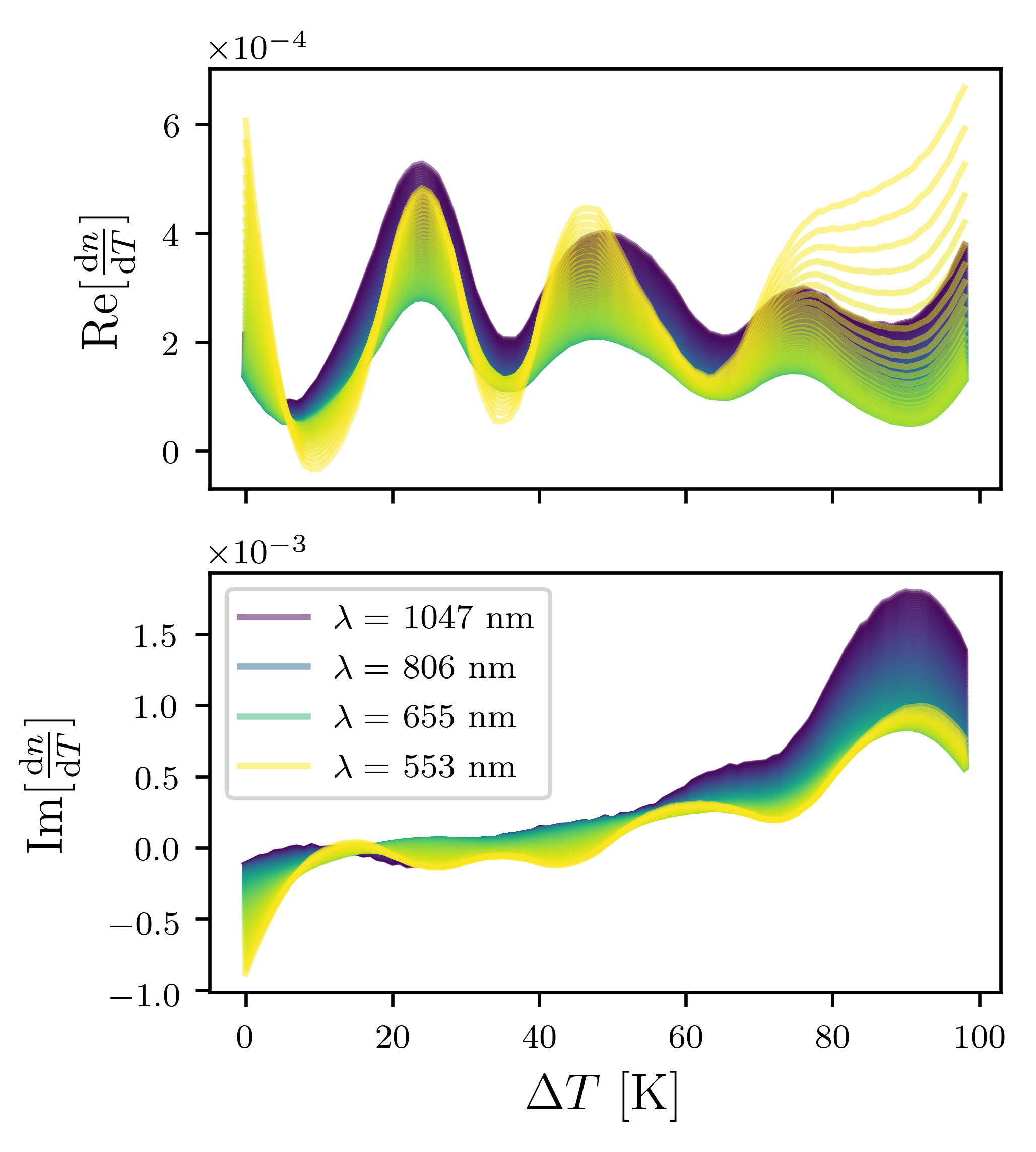} 
\caption{ \label{fig:metal_dndT}
    Modeled change in gold refractive index with temperature from Drude-Lorentz fit to data \cite{zahra} at each temperature independently. The frequency dependence is nontrivial, as the imaginary part especially varies from $\ll 10^{-4}$ 1/K (typical of solvents) to $> 10^{-3}$.
    }
\end{figure}

For the linear polarizability model, the real and imaginary parts of $\frac{dn}{dT}$ were determined by finite difference from the modeled refractive index. These values have surprising temperature and frequency dependence, which is plotted in Fig.\ \ref{fig:metal_dndT}.

\begin{table}[!ht]
\centering
\begin{tabular}{ |c||c|c|c|c||c| } 
    \hline
    Medium & $n$ & $\frac{\mathrm{d}n}{\mathrm{d}T}$ & $C_p$ [J/m$^3$K] & $\kappa$ [W/mK]& SNR, Ref.~\cite{gaiduk2010detection}
    \\ 
    \hline\hline
    water \cite{gaiduk2010detection} & 1.33 & $-9\times 10^{-5}$& $4.2 \times 10^6$ & 0.6 & 0.133
    \\
    \hline
    glycerol \cite{gaiduk2010detection} & 1.473 & -2.70$\times 10^{-4}$& 2.60$\times10^5$ & 0.292 & 0.989
    \\
    \hline
    ethanol \cite{gaiduk2010detection} & 1.36 & -4.40$\times10^{-4}$ & 1.93$\times10^{6}$ & 0.167 & 1.994
    \\
    \hline
    hexane \cite{de2001temperature} & 1.37 & -5.50$\times10^{-4}$& 1.50$\times10^{6}$ & 0.124 & 4.008
    \\
    \hline
    pentane \cite{gaiduk2010detection} & 1.358 & -5.99$\times10^{-4}$& 1.45$\times10^{6}$ & 0.111 & 5.056
    \\
    \hline
    chloroform \cite{gaiduk2010detection} & 1.45 & -6.19$\times10^{-4}$ & 1.434$\times10^{6}$ & 0.13 & 2.839
    \\
    \hline
    dichloromethane \cite{valkai1998temperature} & 1.422 & -7.4$\times10^{-4}$ & 1.576$\times10^{6}$ & 0.1392 & 3.3
    \\
    \hline
    5CB$_e$ \cite{ono2003simple} & 1.72 & -3.01$\times10^{-3}$ & 2.00$\times10^6$ & 0.16 & n/a
    \\
    \hline
    5CB$_o$ \cite{ono2003simple} & 1.54 & 6.37$\times10^{-4}$ & 2.00$\times10^6$ & 0.16 & n/a
    \\
    \hline
\end{tabular}
\caption{
    Thermal and optical properties for background media. Most values for the change in refractive index with temperature are from Refs.\ \cite{gaiduk2010detection,chang2012enhancing}.
}
\label{table:1}
\end{table}

\section{Effect of the substrate}\label{section:trans_refl_figs}

\begin{figure} 
\includegraphics{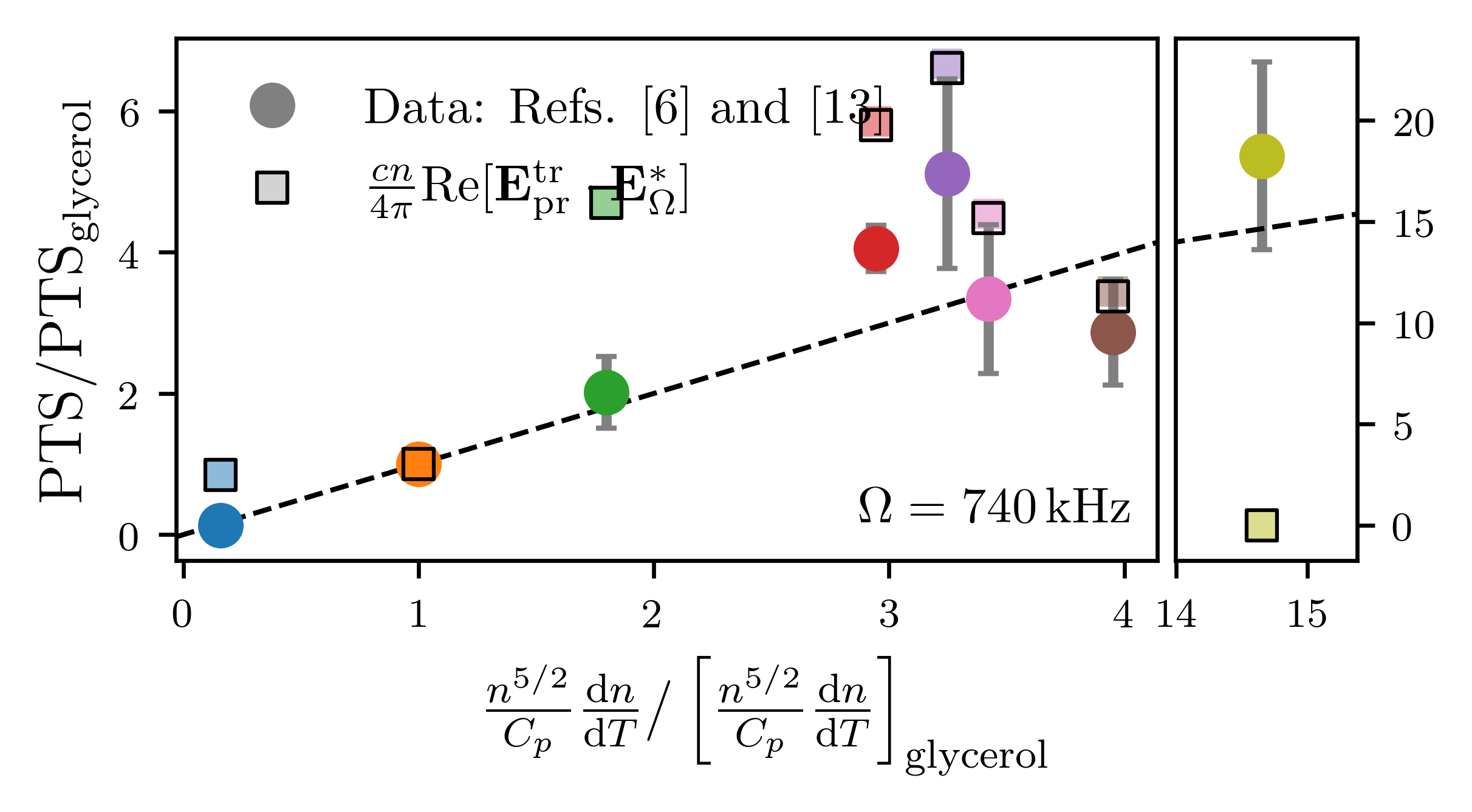} 
\caption{ \label{fig:1_w_glasskappa_reflGBG}
    Including the qualitative effects of reflection at the glass/background/glass interfaces as well as the glass substrate on thermal diffusion. The reflection was done neglecting interference effects between reflected rays at each interface. The glass correction to the temperature was done by naively averaging the glass thermal conductivity with the background and treating the sphere as still in isotropic medium. 
    }
\end{figure}

\begin{figure} 
\includegraphics{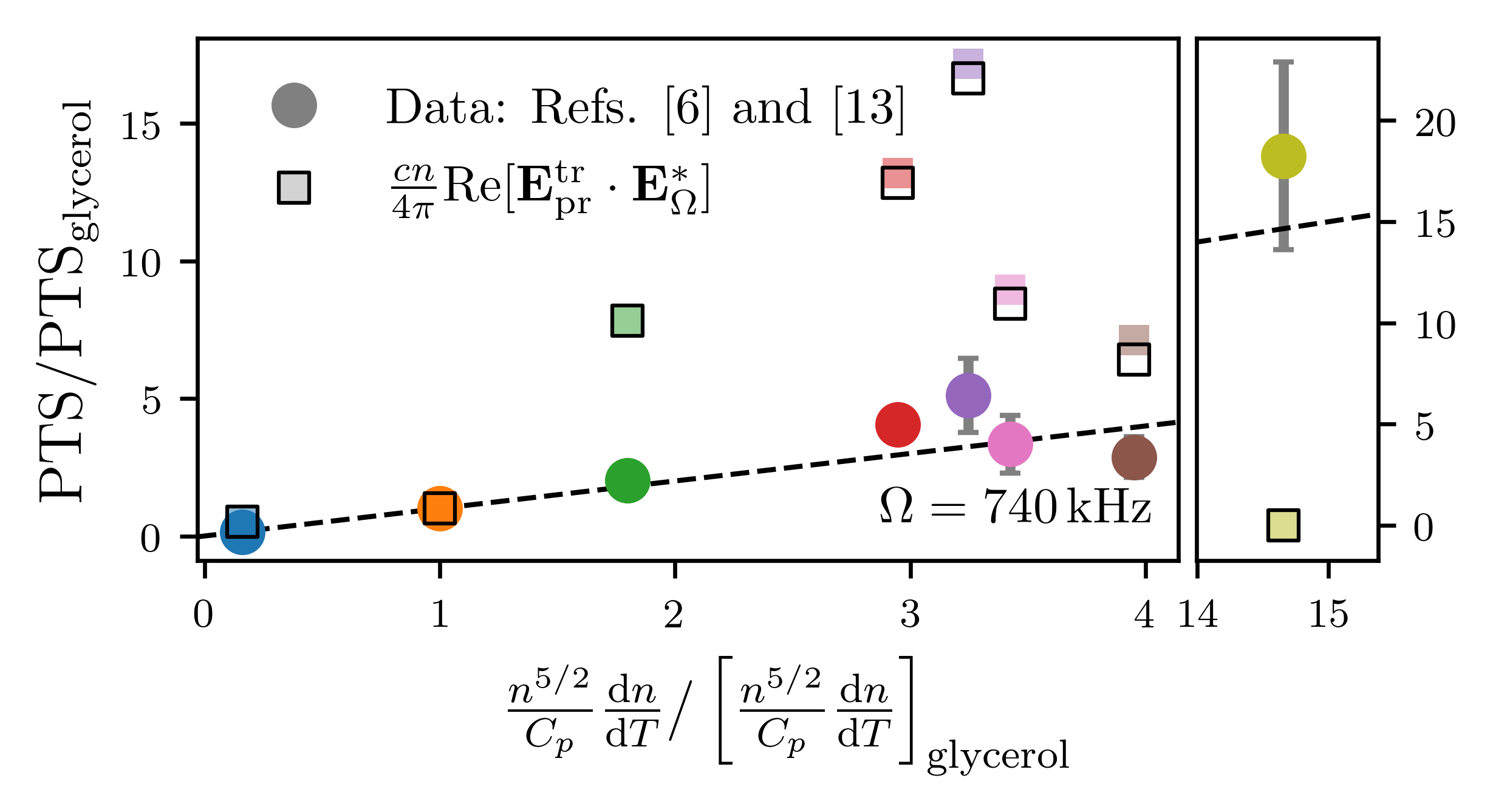} 
\caption{ \label{fig:1_w_reflGBG}
    Including the qualitative effect of the glass/background/glass interfaces on the reflection coefficient (but not the thermal effect of the substrate). In this case the self-interference term contributes non-negligibly. The probe-interference term is therefore shown by the shaded+colored squares and the total photothermal signal is shown in the square outlines.
    }
\end{figure}

\begin{figure} 
\includegraphics{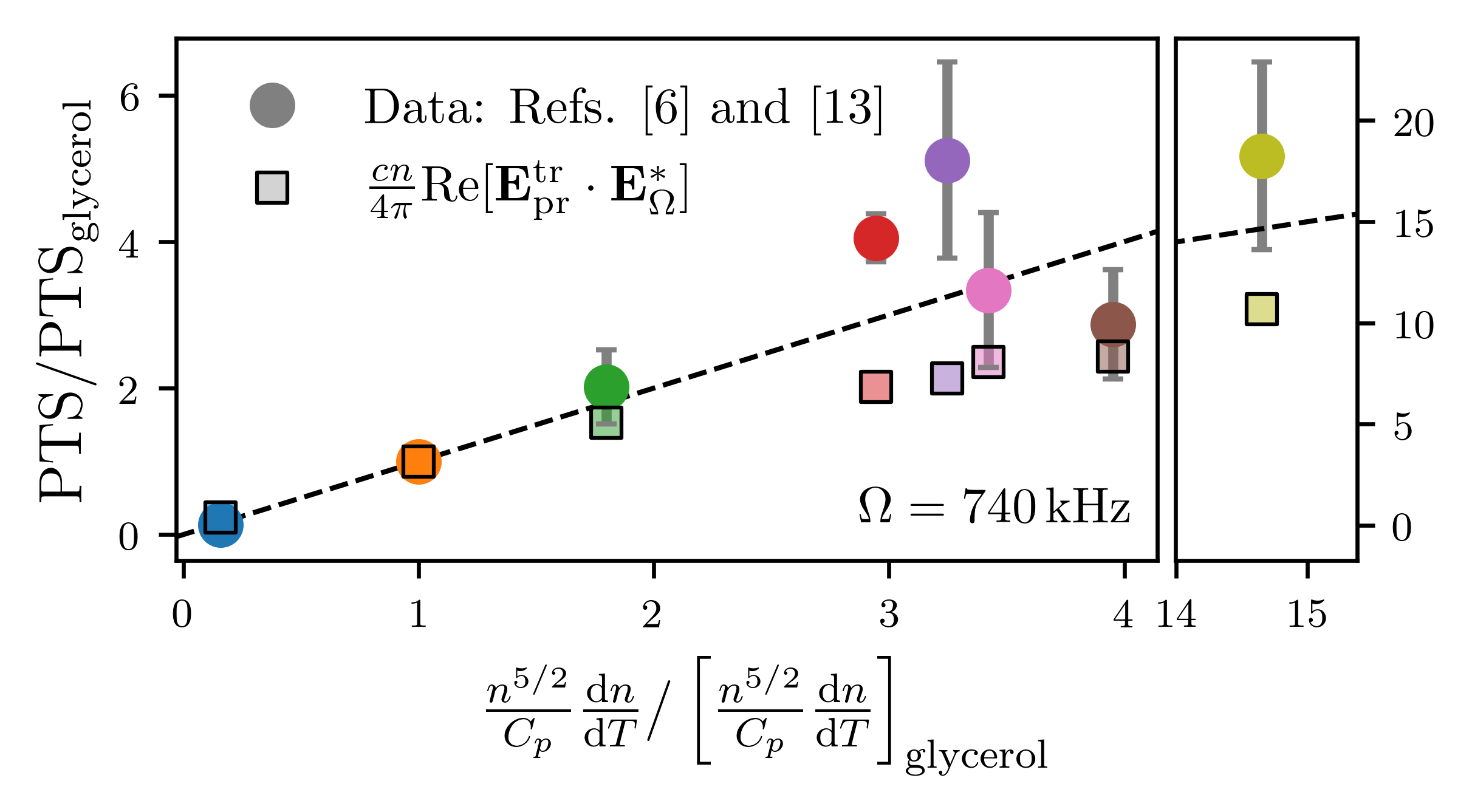} 
\caption{ \label{fig:1_w_glasskappa}
    Including the qualitative effect of the glass substrate as a heat sink (but not the reflective effect of the substrate). This was done naively by averaging the thermal conductivity for glass and background and treating the sphere as still in isotropic medium.
    }
\end{figure}

Figs.\ \ref{fig:1_w_glasskappa_reflGBG}, \ref{fig:1_w_reflGBG}, and \ref{fig:1_w_glasskappa} illustrate a qualitative model of the effects of the substrate on the trend in photothermal signal with background medium presented in main text Fig.\ 2. The substrate should impact the measurement in two primary ways not accounted for by the model. First, the data from Ref.\ \cite{gaiduk2010detection} 
was acquired in the reflection geometry, so the refractive index change at the substrate/background interface should affect the photothermal signal. To explore this, the reflection coefficient at the various interfaces involved was calculated neglecting the field phase. The second effect is on heat diffusion due to the increased thermal conductivity of glass compared to the solvents assumed to be isotropic background in the model. This case was explored here by naively averaging the thermal conductivity of glass and background. Combining there two effects (Fig. \ref{fig:1_w_glasskappa_reflGBG}) qualitatively reproduces the trend in the data arguably better then when neglecting these effects in the main text. But the qualitative nature of these additions clearly has problems, with a noticeable offset for ethanol, hexane, and pentane, as well as the prediction for 5CB trending to zero. The work presented here is taken as evidence that a more quantitative model of substrate effects may recover the experimental trend in the data precisely. 

\begin{figure} 
\includegraphics[width=\textwidth]{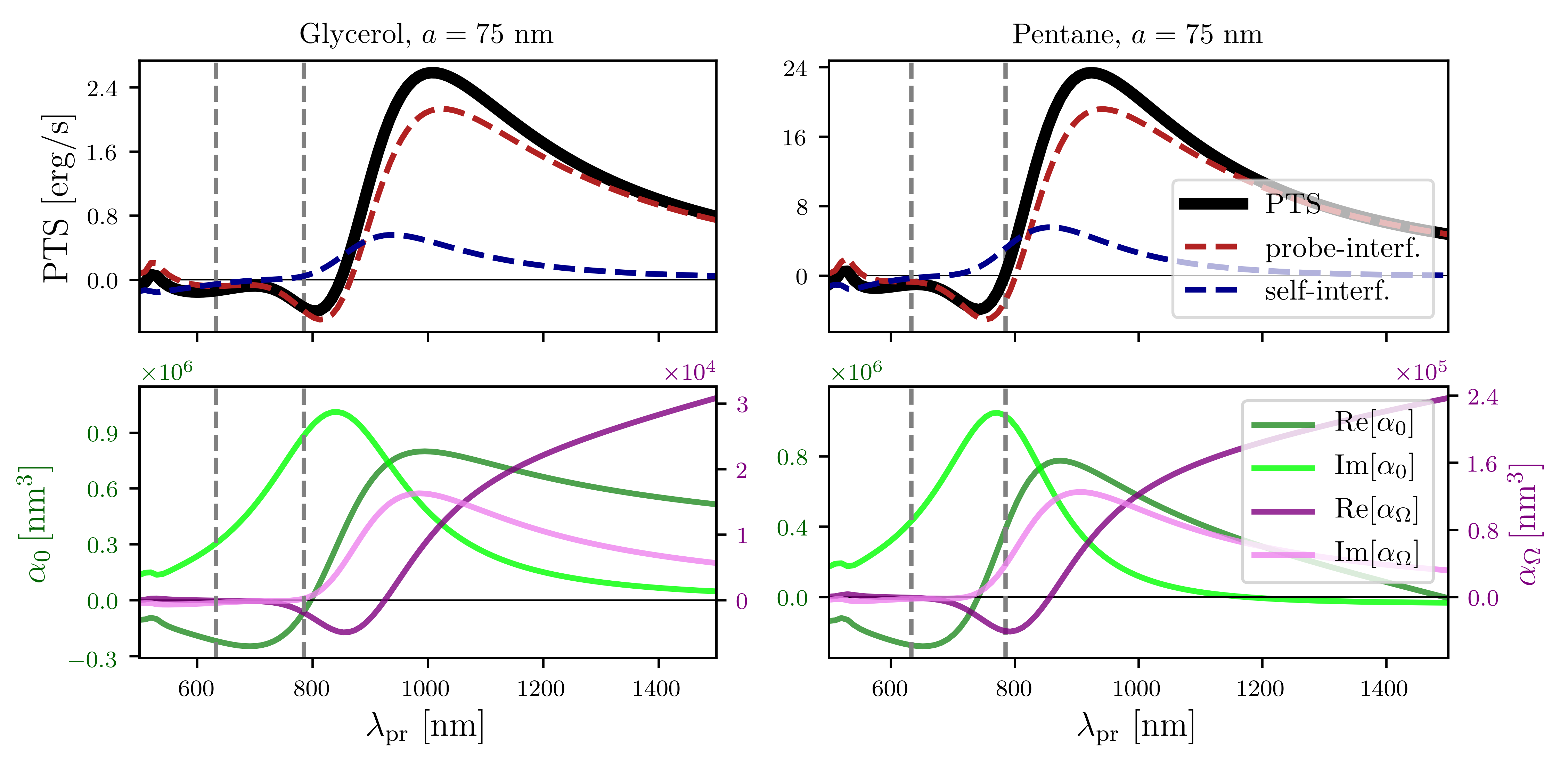} 
\caption{ \label{fig:4_highNAcol}
    Photothermal signal (top) and polarizability components for larger $a=75$ nm particles at $\mathrm{NA}_\mathrm{col} = 1.4$ compared to $\mathrm{NA}_\mathrm{col} = 0.7$ in main text Fig.\ 5. The increased self-interference contribution (blue) demonstrates an example of tuning the probe- and self-interference by numerical aperture of the collection objective. Increasing the $\mathrm{NA}_\mathrm{col}$ generates more photothermal signal, but here the pump and probe powers have been kept fixed. The transmitted/reflected probe field therefore increases less relative to the other fields, and the self-interference gains on the probe-interference.
    }
\end{figure}
\clearpage

\section{Alternate interpretation of the modulated pump and resulting lock-in photothermal signal}\label{section:mod_E}

Before embarking on an alternative ansatz for the form of the pump beam, we can examine the form chosen for the main text a bit more. Although the expression given in Eq.\ (4) for the modulated pump intensity was an ansatz, we may derive it rigorously from a real pump electric field of the form
\begin{align}
\mathbf{E}_\mathrm{pu}(t) 
={}& 
\mathrm{Re}[
    \mathbf{E}_\mathrm{pu} e^{-i\omega_\mathrm{pu}{t}}
    \cdot\tfrac{1 + e^{-i\Omega{t}}}{\sqrt{2}}
    ]
\\
={}& 
\frac{\mathbf{E}_\mathrm{pu}}{\sqrt{2}} \left[
    \cos\omega_\mathrm{pu}{t}
    +
    \cos(\omega_\mathrm{pu} + \Omega){t}
    \right]
\end{align}
Computing the time average explicitly, 
\begin{align}
{I}_\mathrm{pump}(t; \omega_\mathrm{pu})
={}&
\frac{c n}{4\pi}
\frac{\omega_\mathrm{pu}}{2\pi}
\int_0^{\frac{2\pi}{\omega_\mathrm{pu}}} E_\mathrm{pu}^2(t)
\\
={}&
\frac{c n}{4\pi}
\frac{\omega_\mathrm{pu}}{2\pi}
\frac{{E}_\mathrm{pu}^2}{2}
\int_0^{\frac{2\pi}{\omega_\mathrm{pu}}}
     \left[
        \cos\omega_\mathrm{pu}{t}
        +
        \cos\omega_\mathrm{pu}t\cos\Omega{t} 
        -
        \sin\omega_\mathrm{pu}t\sin\Omega{t}
        \right]^2
\;.\end{align}  
Taking the relevant timescale limit mentioned in the submitted manuscript: $\Omega \ll \omega_\mathrm{pu}$, we can consider $\Omega{t}$ to be time-independent for the integral domain. The above then becomes 
\begin{align}
{I}_\mathrm{pump}(t; \omega_\mathrm{pu})
={}&
\frac{c n}{4\pi}
\frac{\omega_\mathrm{pu}}{2\pi}
\frac{{E}_\mathrm{pu}^2}{2}
\left[
    ( 1 + \cos\Omega{t} )^2
    \int_0^{\frac{2\pi}{\omega_\mathrm{pu}}}
    \cos^2\omega_\mathrm{pu}t
    +
    \sin^2\Omega{t}
    \int_0^{\frac{2\pi}{\omega_\mathrm{pu}}}
    \sin^2\omega_\mathrm{pu}t
        \right]
\;,\end{align}  
plus two cross terms that integrate to zero. Both integrals inside the square brackets equal compute to $\frac{\pi}{\omega_\mathrm{pu}}$, which cancels with the overall prefactor. After some manipulation of the trigonometric expressions, we arrive at the result
\begin{align}
{I}_\mathrm{pump}(t; \omega_\mathrm{pu})
={}&
\frac{c n}{8\pi}
{E}_\mathrm{pu}^2
\frac{1+\cos\Omega{t}}{2} 
\;.\end{align}  
                                                   
If we instead begin the derivation of the lock-in photothermal signal not with an intensity modulated pump beam, but instead with the pump electric field amplitude modulated, 
\begin{equation}
\mathbf{E}^\prime_\mathrm{pu}(t) 
= 
\mathbf{E}_\mathrm{pu} \cos\omega_\mathrm{pu}t \,\frac{1+\cos\Omega{t}}{2}
\;,
\end{equation} 
written in real form. 
We will show here that this starting assumption leads to the same decomposition of the lock-in photothermal signal into probe- and self-interference components discussed in the main text. Rather surprisingly, this alternative pump field also leads to a photothermal signal that reduces into the same thermally-static limit as stated in the main text. 

Assuming $\Omega \ll \omega_\mathrm{pu}$, the slowly varying pump intensity resulting form the real electric field written above is
\begin{equation} \label{eq:I_pu_prime}
{I}^\prime_\mathrm{pump}(t; \omega_\mathrm{pu})
={}
\frac{c n}{8\pi}
\left|\mathbf{E}_\mathrm{pu} \right|^2
\, \frac{1 + 2\cos\Omega{t} + \cos^2\Omega{t}}{4}
\;.
\end{equation}
Summarizing the analogous derivation in the main text, the linear pump absorption and linear model of the temperature dependent polarizability leads to a scattered field of the form,
\begin{equation} \label{eq:Escat_prime}
\mathbf{E}^\prime_\mathrm{scatt}(t)  
= 
\mathbf{E}_0^\prime 
+
\mathbf{E}_\Omega^\prime \cos\Omega{t}
+
\mathbf{E}_{\Omega2}^\prime \cos^2\Omega{t}
\;.\end{equation}
Just like in the main text, the average of the time dependent portion of Eq.\ \ref{eq:I_pu_prime} is $1/2$, and the nanoparticle temperature can be written in the compact form, 
\begin{equation}
\Delta T_\mathrm{NP}(t)
= 
\langle{T_\mathrm{NP}}\rangle
\left(
    \frac{1}{2} + \cos\Omega{t} + \frac{1}{2}\cos^2\Omega{t}
    \right)
\;.\end{equation}
Underlying Eq.\ \ref{eq:Escat_prime} for the scattered field $\mathbf{E}^\prime_\mathrm{scatt}(t)$ is linear polarizability $\boldsymbol{\alpha}(t) = \boldsymbol{\alpha}_0 + \boldsymbol{\alpha}_\Omega\cos\Omega{t} +\boldsymbol{\alpha}_{\Omega2}\cos^2\Omega{t}$ defined by the 3 terms, 
\begin{align}
\boldsymbol{\alpha}_0 
={}& 
    \boldsymbol{\alpha}({T_R})
    + 
    \frac{{d}{\boldsymbol{\alpha}}}{{d}n} \frac{{d}n}{{d}T}
    \langle{T_\mathrm{NP}}\rangle\frac{1}{2}
\\
\boldsymbol{\alpha}_\Omega
={}&  
    \frac{{d}{\boldsymbol{\alpha}}}{{d}n} \frac{{d}n}{{d}T}
    \langle{T_\mathrm{NP}}\rangle
\\
\boldsymbol{\alpha}_{\Omega2}
={}&  
    \frac{{d}{\boldsymbol{\alpha}}}{{d}n} \frac{{d}n}{{d}T}
    \langle{T_\mathrm{NP}}\rangle\frac{1}{2}
\;,\end{align}
where derivatives can be expanded in the core shell model as $\frac{{d}{\boldsymbol{\alpha}}}{{d}n}\frac{{d}n}{{d}T} = \frac{{d}{\alpha}}{{d}n_c} \frac{{d}n_c }{{d}T} + \frac{{d}{\alpha}}{{d}n_s}\frac{{d}n_s }{{d}T} $ without loss of generality. 

Computing the lock-in integral over the raw photothermal signal generates a new term, containing the interference between the scattered field of order $\cos\Omega{t}$ and the higher harmonic of order $\cos^2\Omega{t}$,
\begin{equation}
\begin{split}
{P^\text{PT}}^\prime 
&\equiv
\lim_{\tau\to\infty} \frac{1}{\tau}\int_0^\tau  {d}t\,  {P_\mathrm{PT}^\mathrm{raw}}^\prime  \cos(\Omega t)\\
&= 
\frac{c n}{4\pi}
\mathrm{Re}\!\left[
    {{E}_\mathrm{pr}^\mathrm{tr}}\cdot {{{E}_\Omega}^\prime}^* 
    + 
    {{{E}_0}^\prime} \cdot {{{E}_\Omega}^\prime}^*  
    +
    {{{E}_\Omega}^\prime} \cdot {{{E}_{\Omega2}}^\prime}^*
    \right]
\frac{f(\theta_\mathrm{col})}{k^2}\\
&\equiv {P^\mathrm{PT}_\mathrm{PI}}^\prime + {P^\mathrm{PT}_\mathrm{SI}}^\prime.
\;,\end{split}
\end{equation} 
and we see that the probe interference term is identical to the approach described in the manuscript while the self-interference has been modified. This expression does simplify by noting that ${E}_{\Omega2} = \tfrac{1}{2}{E}_{\Omega}$, and the self-interference can be written
\begin{equation}
{P^\mathrm{PT}_\mathrm{SI}}^\prime 
=
\frac{c n}{4\pi}
\mathrm{Re}\!\left[
({{{E}_0}^\prime} + \tfrac{1}{2}{{E}_\Omega}^\prime) \cdot {{{E}_\Omega}^\prime}^*  
\right]
\frac{f(\theta_\mathrm{col})}{k^2} 
\;. 
\end{equation}
Surprisingly, this reduces to the same thermally static limit as in the main text. Here we note that the static hot and cold fields can be defined 
\begin{align}
E_H ={}& E_0 + \tfrac{3}{2}E_\Omega
\\
E_C ={}& E_0 - \tfrac{1}{2}E_\Omega
\end{align}
and therefore
\begin{align}
E_\Omega ={}& \frac{E_H - E_C}{2}
\\
E_0 ={}& \frac{E_H + 3E_C}{4}
\end{align}
from this we find that 
\begin{equation}
\mathrm{Re}\!\left[
({{{E}_0}^\prime} + \tfrac{1}{2}{{E}_\Omega}^\prime) \cdot {{{E}_\Omega}^\prime}^*  
\right]
=
\tfrac{1}{4}[|E_H|^2 - |E_C|^2]
\;.\end{equation}

\section{Signal dependence upon the particle–probe-focus offset}
\begin{figure} 
\includegraphics[width=\textwidth]{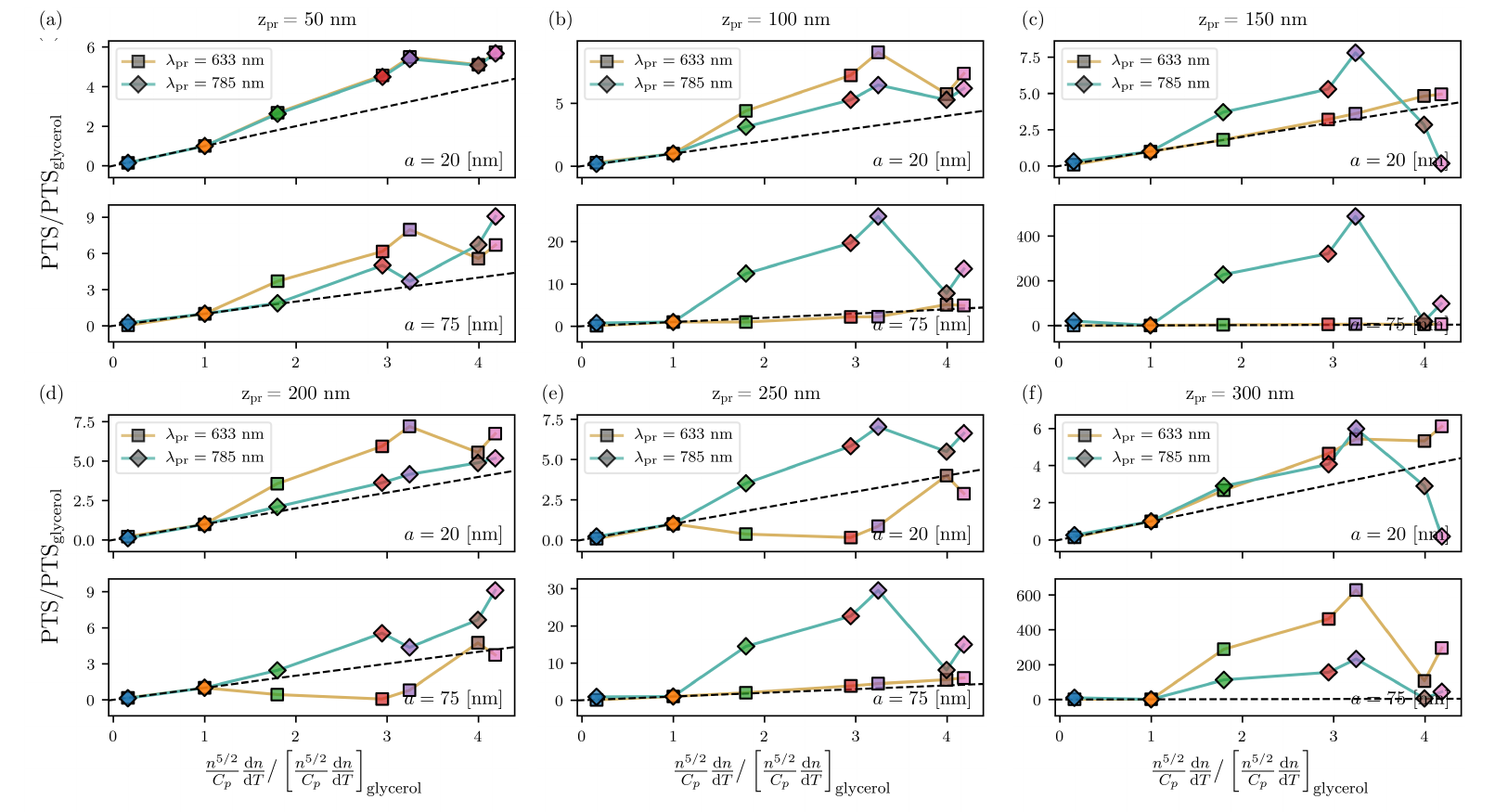} 
\caption{Display of the drastic affect of the particle--probe-focus offset $z_\mathrm{pr}$ on the trend in photothermal signal with background medium presented in the main text Fig.\ 4. 
\label{new}}
\end{figure}

Fig. \ref{new} displays the dependence of the photothermal signal upon the particle–probe-focus offset at different probe wavelengths and in different solvents.
\clearpage

\bibliography{refs}